\documentclass[NETN]{stjour}
\articletype{Research}
\usepackage{rotating}
\usepackage{tabularx}
\usepackage{booktabs}
\usepackage[english]{babel}
\usepackage{varioref}
\usepackage{hyperref}
\usepackage{cleveref}
\usepackage{changepage}
\usepackage{csquotes,booktabs,dcolumn}
\usepackage[title]{appendix}
\usepackage{textpos}
\usepackage{graphicx}%
\usepackage{dcolumn} %
\usepackage{bm}      %
\usepackage{subfigure}
\usepackage{amsfonts}
\usepackage{rotating}
\usepackage{fontenc}
\usepackage{cancel}
\usepackage{amsthm}
\usepackage{hyperref}
\usepackage[normalem]{ulem}
\usepackage{longtable} 
\usepackage{adjustbox} 
\usepackage{array}
\usepackage{booktabs}
\usepackage{multirow}

\newcolumntype{d}[1]{D{.}{.}{#1}}
\newcolumntype{Y}{>{\centering\arraybackslash}X}

\begin{document}
\title[The dominance of big teams in China's scientific output]{The dominance of big teams in China's scientific output}

\author[Author Names]
{Linlin Liu\affil{1},
Jianfei Yu\affil{1}, Junming Huang\affil{2,3},Feng Xia\affil{4}\\
\and Tao Jia\affil{1,5*}}

\affiliation{1}{College of Computer and Information Science, Southwest University, Chongqing, 400715, P. R. China.}

\affiliation{2}{Paul and Marcia Wythes Center on Contemporary China, Princeton Institute for International and Regional Studies, Princeton University, Princeton, NJ 08540, USA}

\affiliation{3}{Center for Complex Network Research, Northeastern University, Boston, Massachusetts 02115, USA}

\affiliation{4}{School of Science, Engineering and Information Technology, Federation University Australia, Ballarat, VIC 3353, Australia}

\affiliation{5}{Deakin-SWU Joint Research Center on Big Data, Southwest University, Chongqing, 400715, P. R. China}

\correspondingauthor{Tao Jia}{tjia@swu.edu.cn}



\begin{abstract}
Modern science is dominated by scientific productions from teams. A recent finding shows that teams with both large and small sizes are essential in research, prompting us to analyze the extent to which a country's scientific work is carried out by big/small teams. Here, using over 26 million publications from Web of Science, we find that China's research output is more dominated by big teams than the rest of the world, which is particularly the case in fields of natural science. Despite the global trend that more papers are done by big teams, China's drop in small team output is much steeper. As teams in China shift from small to large size, the team diversity that is essential for innovative works does not increase as much as that in other countries. Using the national average as the baseline, we find that the National Natural Science Foundation of China (NSFC) supports fewer small team works than the National Science Foundation of U.S. (NSF) does, implying that big teams are more preferred by grant agencies in China. Our finding provides new insights into the concern of originality and innovation in China, which urges a need to balance small and big teams.
\end{abstract}


\section{Introduction}
Modern science has witnessed the increasing dominance of teams. The single‐author papers, though not yet distinct as what Price predicted in 1963 \citep{de1963little}, have undergone a sharp drop, taking only a small portion of all publications \citep{wuchty2007increasing,lariviere2015team,barlow2018extinction}. Teams become the driving force of science because not only the problem to tackle is more complex, but also the knowledge is broader, which inevitably makes scientists more specialized \citep{jones2009burden, leahey2016sole}. The improvement of communication technology, the convenience of transportation and the globalization also facilitate scientific collaborations. All of these make teams not only flourishing but also grow by size \citep{newman2001scientific, gazni2012mapping, lariviere2015team,wu2019large}. The average number of authors per publication increases every year and large teams involving more than 1000 members become common. In a recent paper studying the mass of the Higgs boson, the team size reaches a record high of over 5,000 scientists\citep{castelvecchi2015physics}.   

The large team has clear advantages over the small team in solving complicated problems, securing research grants, receiving more citations on average, and publishing hit papers that are on the top of citation rank \citep{thelwall2019large, wuchty2007increasing, cummings2007coordination}. Recent research shows, however, that the bigger is not always the better \citep{wu2019large}. Instead, small and large teams take distinct yet equally essential roles in science. Large teams tend to work in the established field and exploit existing problems. In contrast, small teams are better at exploring the frontier of science, generating new ideas, and opening up new problems that can disrupt science. To better promote science, a balance between small and large teams is needed \citep{azoulay2019small}, giving rise to an interesting question: to what extent the research works of a nation is carried out by big/small teams.

The answer to this question may have important implications to the scientific performance of a nation, if we accept the fact that patterns observed in small and large teams are universal. While it is hard to argue whether a balance or an optimal is reached in a nation, it is still meaningful to compare the small/large team composition in different countries. This is of particular importance to China in the context of its long term goal to be a global innovator \citep{phillips2016china, zhou2016china}. Indeed, while China has grown to be the world's top scientific paper producer and citation receiver, it is often worried that China's scientific works are weak in originality and innovation \citep{xie2014china, huang2018quality, guo2019contributions}. In this paper, we perform quantitative analyses on over 26 million papers published from 2000 to 2017. We find that China is indeed different from other countries. The percentage of China's scientific annual output from small teams is now the lowest in the world, after a sharp drop since 2000. As research teams in China shift from small to large size, the team diversity that is essential for innovative works does not increase as much as that in other countries. Most works by big teams are still carried out in one or two institutes. The dominance of big teams in China may not be explained by the citation boost from the team size. While the citation on average increases with team size, the rate of increase is roughly the same in every country. However, the preference of funding agencies may be related to the lack of small teams in China. In all, if small teams are more apt to perform disruptive research, the science community in China should be alerted, given the different statistics China demonstrates.

\section{Data and Methods}
{\bf Data set.} We use the publication data of the Web of Science (WOS), covering the Science Citation Index Expanded (SCIE), Social Sciences Citation Index (SSCI) and Arts \& Humanities Citation Index (A \& HCI) databases. In total, there are over 26 million publications from year 2000 to 2017, including 18,295,191 articles, 3,646,465 meeting abstracts, 1,255,019 proceedings papers, 1,055,520 reviews, 970,649 editorial materials, 600,187 letters, 447,620 book reviews, 79,121 corrections, 32,205 biographical items, 29,115 news items, and more. The variety of document types naturally prompts to check if the conclusion would change when a different set of documents are considered. In particular, we perform a separate analysis by considering more ``traditional'' form of scientific papers, including article, review, letter and conference proceeding. We find that there are only small changes on the statistics and conclusions we drawn remain the same.

{\bf Statistical Test.} The statistical test is crucial when the sample size is relatively small. The p-value is usually reported to gauge if the difference between the two measures is statistically significant. However, given the size of the data applied in our study, we find that most of our comparisons are statistically significant ($p \le 0.05$). This can be demonstrated in a theoretical manner. Consider a very general case in our analysis: there are two proportions $p_1$ and $p_2$ based on two samples with size $n_1$ and $n_2$, respectively. To test if $p_1$ is significantly bigger or smaller than $p_2$, we need to use the one-tail z-test. To simplify the model a bit more, let's consider the smaller one in $p_1$ and $p_2$ is $p_o$ and the larger one is $p_o+\delta$. The two samples can be approximated with equal size that $n_1 = n_2 = n$. We can then plug the parameter $p_o$, $\delta$ and $n$ into the calculation of the z-score (which consequently gives the p-value). When $n = 15,000$, the difference $\delta = 0.01$ is guaranteed to be statistically significant regardless of the $p_o$. Because most samples in this study are with size greater than 15,000, it means that almost any virtual difference in the figure is statistically significant. Therefore, we choose not to report the p-value repetitively. If not otherwise mentioned, two proportions are statistically different. Indeed, there is only one instance (which is explicitly mentioned) in the paper, where the two measures are so close that the difference is not significant.

{\bf Country Allocation.} We use the straight counting by the first affiliation \citep{waltman2015field, zheng2014influences, huang2011counting}. The country of a paper's first affiliation determines the country this paper belongs to. Other methods such as whole counting and fractional counting \citep{sivertsen2019measuring, lewison2010understanding, lin2013influences, kao2009authorship,larsen2008state} are also widely applied to count publication numbers. But they may bring the issue of multiple counting, which can be a problem in the analysis. Previous works suggest that straight counting might be better when studying the scientific output at the country level \citep{huang2011counting}. 
Another strategy of straight counting is to use the corresponding affiliation or the so-called ``reprint address'' in WOS database \citep{kahn2016return, mazloumian2013global,gonzalez2017dominance}. We find that for more than 95\% of papers, the reprint address and the first affiliation point to the same country. For simplicity and the ease of future reproduction of our analyses, we choose to use the first affiliation, as the information of corresponding affiliation may not be directly available in other databases.
Finally, to eliminate possible bias caused by straight counting in dealing with papers by international collaborations, we also separately analyze publications by authors from the same country. We find that our conclusion is not affected (Supplementary Note 1).

{\bf Countries considered.} We include 15 countries in our analyses, which are roughly the top 15 countries of total scientific publications in our analysis (except for Turkey which ranks on the 16th). They are United States (US), China (CN), United Kingdom (GB), Germany (DE), Japan (JP), Italy (IT), France (FR), Canada (CA), India (IN), Korea (KR), Spain (ES), Australia (AU), Brazil (BR), Netherlands (NL) and Turkey (TR). Given China's huge annual production of scientific papers, it is less meaningful to compare it with countries of less scientific output. Following the typical practices, we use the scientific production from the mainland of China, Hong Kong, China and Macau, China. We remove China when taking a global average in order to show a more vivid comparison between China and the rest of the world.

{\bf Big team and small team.} The term big team and small team are relatively new and there is no defined hard cutoff between them. Previous work \citep{wu2019large} considers team size ($m$) of no more than 3 or 4 members as small. In this work, we analyze all situations ($m \le 3$, $m \le 4$, $m \le 5$) and find that our conclusion in general is not affected by the choice of parameters. The only inconsistency is that $P(m \le 5)$ of China is slightly higher than that of Japan and Italy, making China not the lowest, but the third lowest among the 15 countries. The value is still way below the global average. We present results based on $m \le 4$ in the main text of the paper. The corresponding results for $m \le 3$ and $m \le 5$ are presented in the Supplementary Information.

{\bf Research field of a paper.} WOS has approximately 250 subject areas characterizing different research directions. Each paper is assigned one or multiple subject areas. The large number of subject areas makes it impossible to draw any conclusions in different research directions. Therefore, we use the classification in \cite{wu2019large} that merges WOS subject areas into 14 research fields, including physical sciences, chemistry, biology, medicine, agriculture, environmental and earth sciences, mathematics, computer and information technology, engineering, social sciences, business and management, law, humanities, and multidisciplinary sciences. This categorization is slightly different from what is recently proposed by \citep{milojevic2020practical}. However, because the observation is mainly in the field of natural science, the classification difference should not affect the results.
Since a paper is usually tagged by multiple subject areas, it may also be labeled by multiple research fields. It is difficult to tell the priority in multiple subject areas, nor could we artificially tell which research field is most close to the content of the paper, we use the whole counting to classify papers into research fields. In general, depending on the publication year, 20\% - 25\% of papers are labeled by multiple research fields.

{\bf Institute diversity.} WOS records the affiliation of each paper. Starting in 2008, it also records the affiliation of each author, i.e., who is affiliated with what affiliations. Therefore, there are two ways to analyze the institute diversity. One is to use a paper's affiliations directly, the other is to use the ``main'' affiliation of each author. Both approaches have pros and cons. The information of a paper's affiliation is easier to extract and is available for all papers in the data set. But giving the trend that more authors are affiliated with multiple institutions \citep{hottenrott2019rise}, directly using such information may overestimate the institute diversity. In some cases we may also have the number of institutes greater than the number of authors. Use an author's ``main'' affiliation seems to be a more reasonable choice, which is also directly applied in the data set of Microsoft Academic Graph \citep{wang2020microsoft, dong2018collaboration}, but determining the primary affiliation out of others might be non-trivial. In this work, we use both methods to analyze institute diversity. If an author has multiple affiliations, we choose the one with the highest rank in the paper. We parse the institute information using the key value ``organization'' in the data, which usually refers to the university and the research lab. We report the results based on the author's affiliation in the main text. The results based on the paper's affiliation can be found in the Supplementary Information. The two approaches give results slightly different in numbers, but the conclusion drawn are the same. We are also aware about the name disambiguation issue in institute names \citep{donner2019comparing}. This should not affect our analyses because we compare institutes in the same paper. It is very unlikely that authors would write one institute in different ways in one single paper.

{\bf Funding information.} WOS starts in 2008 to record the text related to funding acknowledgment of each paper, from where funding information including the grant agency and grant ID is parsed. Despite concerns on the completeness and accuracy of the data \citep{alvarez2017funding, paul2016characterization, tang2017funding}, it remains to be one of the largest available. We use such information directly as the criteria if a paper is funded. Because our measure is controlled by the national average value, we believe flaws in the data recording should not affect the conclusion.

It is more complicated to search which paper is supported by the National Natural Science Foundation of China (NSFC) or the National Science Foundation (NSF) of the United States, because scientists acknowledge these funding agencies in different ways. For NSFC, the most frequently used name is ``National Natural Science Foundation of China'', but other forms of name such as ``Natural Science Foundation of China'', ``NSFC'', ``National Science Foundation of China'', ``National Nature Science Foundation of China'' and ``National Natural Science Foundation'' are also widely used. The name variations of NSF include ``National Science Foundation'', ``NSF'' and ``National Science Foundation (NSF)''. WOS has performed its own grant name disambiguation (which is available online), but such information is not available in our data set. Therefore, we extract the name of grant agency in each paper from China and the United States, filter out these appearing fewer than 1000 times in the data, and manually identify names associated with NSFC and NSF. These names are list in Table S1 of the Supplementary Information. Other statistics given by our approach are listed in Table S2, which are in line with previous findings \citep{huang2016does, wang2012science}. 

It is noteworthy that the Ministry of Science and Technology (MOST) of China has its own research grants such as the National Basic Research Program of China (973 Program), the National High Technology Research and Development Program of China (863 Program), and the National Key Technology R\&D Program of China. The aim of these grants is to support big research groups (Fig. S13). While they cover a relatively small fraction of scientific papers, the overlap with the NSFC is large. \textcolor{blue}{On average,} around 17.5\% of NSFC supported papers are also supported by MOST, or equivalently 73.3\% of MOST supported papers are simultaneously supported by NSFC. To avoid potential bias, \textcolor{blue}{we focus on those papers ``primarily'' supported by NSFC. 17.5\% of NSFC supported papers are excluded in our analysis of NSFC sponsorship, as they are are also supported by MOST}. Note that the National Institutes of Health of the United States (NIH) also tends to support big groups (Fig. S13). To make the comparison equal, we only consider papers ``primarily'' supported by NSF by ignoring roughly 10.5\% of NSF supported papers that are also supported by NIH. More statistics can be found in Table S3 of the Supplementary Information.

\section{Results}

\begin{figure}[htp]
\centering
\includegraphics[width=1.00\textwidth]{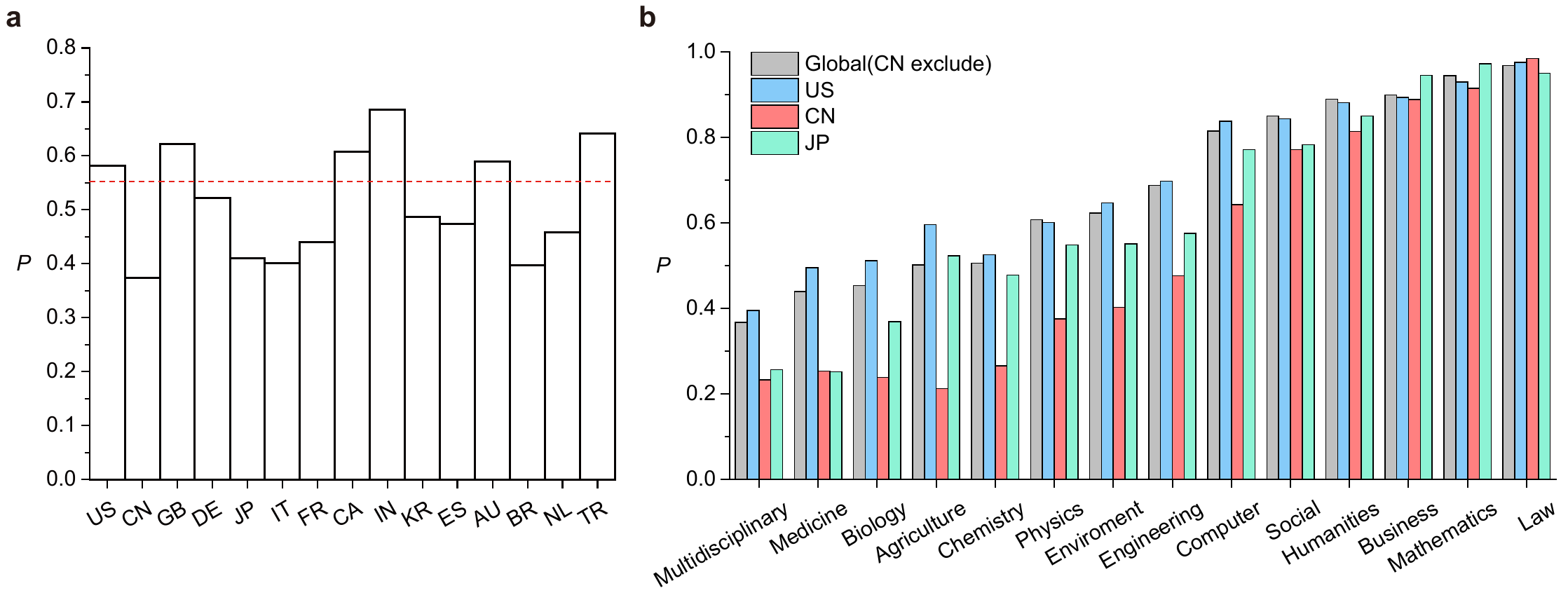}
\caption{{\bf (a)} The fraction of papers in 2017 produced by teams with size no more than 4 ($P(m \le 4)$) in different countries. The dashed line corresponds to the global average (in which China is excluded) {\bf (b)} $P(m \le 4)$ in different fields in the year 2017.}
\label{fig:fig1}
\end{figure}

While collaboration plays an increasingly important role in scientific research, the big team has not yet taken over. In 2017, more than half of scientific papers are produced by teams with relatively small sizes (number of authors $m \le 4$). The fraction of small team output differs from nation to nation, but China ranks the last among the top 15 countries of scientific papers (Fig. 1a and S1). In 2017, only 37\% of papers from China are done by teams with $m \le 4$, while this value is 58\% for United States and 55\% for the global average (in which China is excluded). 

We further analyze the $P(m \le 4)$ in different research fields (Fig. 1b, S1, S2 and S3). The statistics at the global level agree well with previous works and also with our intuitions. For example, small teams are more frequently observed in mathematics, computer science, social science, business, humanities and laws, with $P(m \le 4)$ goes beyond 80\% or even higher. In interdisciplinary fields where collective intelligence is more important, $P(m \le 4)$ drops to the lowest. Fields such as medicine, biology, chemistry, physics, agriculture are usually believed to be labor intensive, requiring more individuals to be involved in. But on the global average, $P(m \le 4)$ is not very far below 50\% and in some fields can even go above. Nevertheless, $P(m \le 4)$ of China is significantly less than the global average in all areas of natural science. The relative difference is most prominent in agriculture, chemistry, biology and medicine. On the contrary, $P(m \le 4)$ of the United States is greater than the global average in almost all areas of natural science. Being an Asian country with large amount of scientific publications, Japan may be expected to be similar as China. But Japan's $P(m \le 4)$ is closer to the global average and is greater than that of China in all areas of natural science except medicine. In fields related to humanity, social science and mathematics, $P(m \le 4)$ of China is not very different from other countries \citep{li2015patterns}, but papers in such fields take only a very small fraction of China's annual production.

\begin{figure}[htp]
\centering
\includegraphics[width=1.00\textwidth]{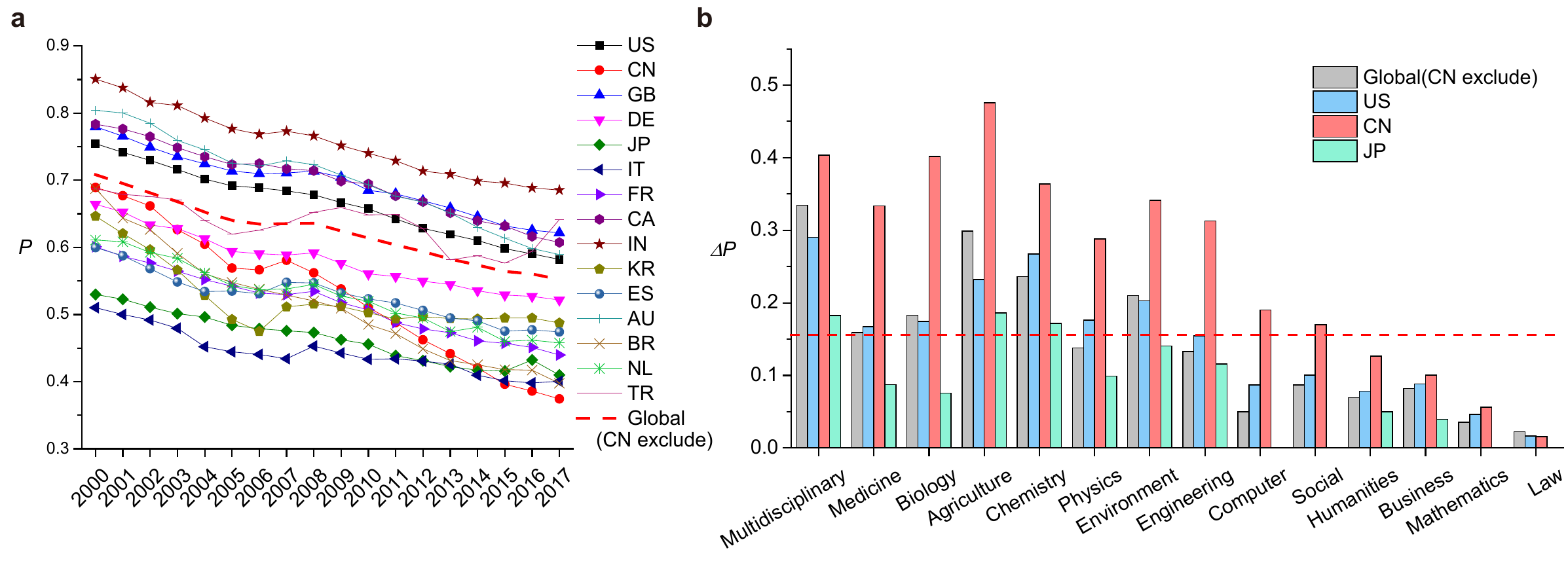}
\caption{{\bf (a)} The time evolution of $P(m \le 4)$ in different countries. The dashed line corresponds to the global average (in which China is excluded). While it is a global trend that more papers are done by big teams, China's drop is much steeper. {\bf (b)} The drop of $P(m \le 4)$ from year 2000 to 2017 in different fields. The dashed line corresponds to the drop of the global value in (a). China's drop is most prominent in fields of natural science and engineering, sometimes can be twice as much as the global value. Note that the small team output slightly increases in Japan in fields of social science and laws, giving rise to a negative value of $\Delta P$. Since we focus on the drop, we do not put them in the figure.}
\label{fig:fig2}
\end{figure}

It is noteworthy that more papers are carried out by big teams is a global trend. Indeed, we find in our analyses that the percentage of papers by small teams decreases over years. Nevertheless, the drop of China is much steeper (Fig. 2a and S4), giving raise to a statistically significant deviation from the global average (Supplementary Note 2). In 2000, $P(m \le 4)$ of China is, though slightly smaller, not very different from that of the United States and global average. But it goes down from 69.9\% in 2000 to 37.4\% in 2017, nearly 32 percentage points decrease. The drop, however, is only 17 percentage points for the United States (from 75.4\% to 58.2\%), 12 percentage points for Japan (from 53.0\% to 41.0\%), and 16 percentage points for the global average (from 70.8\% to 55.2\%). China's drop of small team output in fields of natural science and engineering is much higher than those of global average (Fig. 2b and S4), in line with our initial finding that small team output is small in these fields.

\begin{figure}[htp]
\centering
\includegraphics[width=1.00\textwidth]{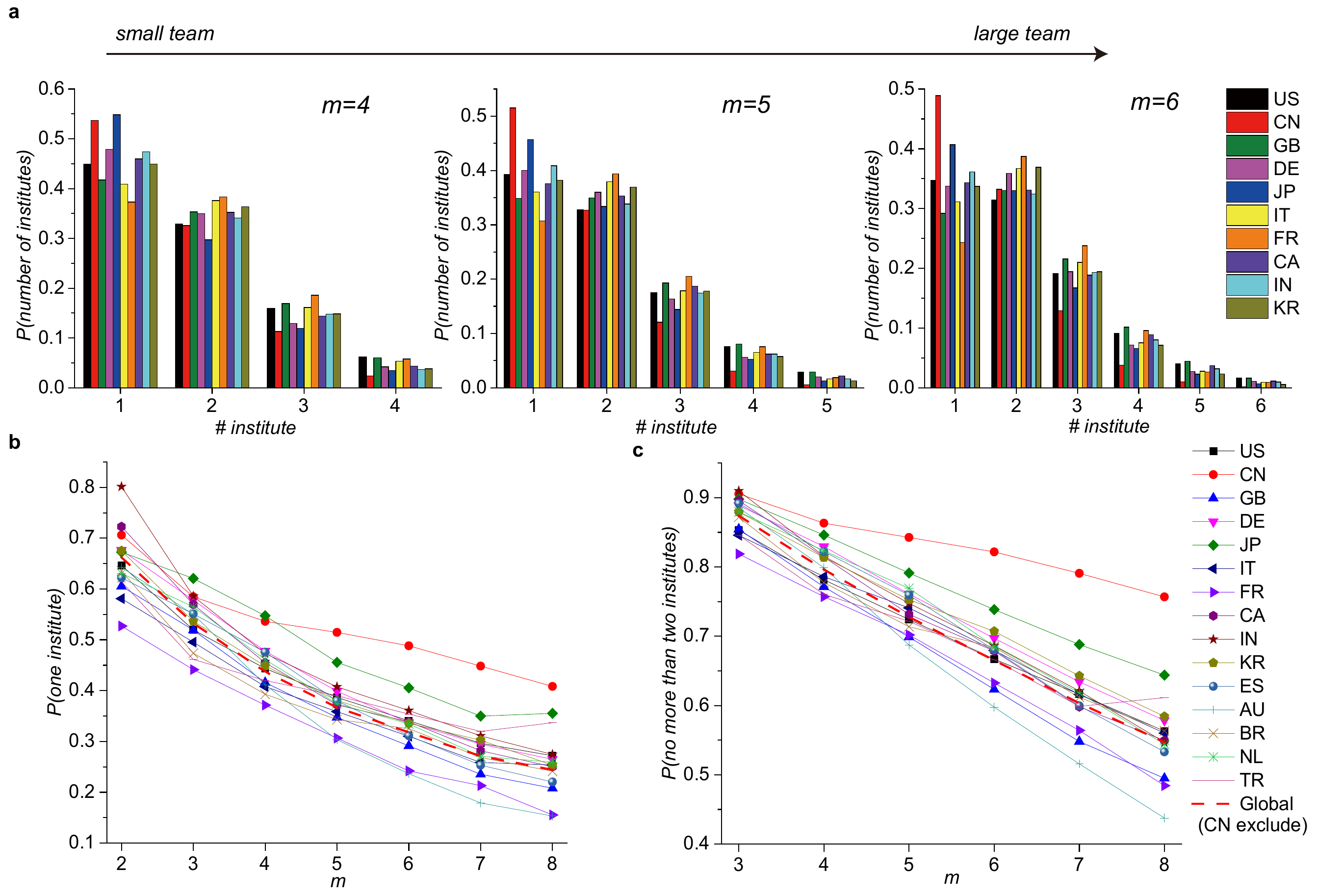}
\caption{{\bf (a)} The distribution of the number of distinct institutes in papers published in 2017 with team size $m = $ 4, 5  and 6 (from left to right). China's team composition is not very different from other countries when the team is small. But as the team size increases, the distribution becomes more dominated by output from one institute. {\bf (b)} The fraction of papers in 2017 done by one institute, given the team size $m$. More percentage of paper output is from a single institute in China than in other countries. {\bf (c)} Similar to (b). The fraction of papers in 2017 involving no more than two institutes. 
}
\label{fig:fig3}
\end{figure}

The observation that big teams dominate China's research output gives rise to another question: how would the team composition change when it shifts from small to large size. Indeed, a team can increase its size by adding more similar members or involving members with different backgrounds. While the team size grows in either way, team diversity is different, which is proved to be an essential factor in building a successful team \citep{alshebli2018preeminence, powell2018these}. There are different types of team diversity, such as ethnicity, discipline, gender, affiliation, and academic age \citep{alshebli2018preeminence, huang2019historical, jia2017quantifying}. Here we focus on the affiliation and analyze the diversity at the institute (organization) level. Indeed, a smaller team whose members are from diverse institutes is more likely to generate ``hit'' papers than a relatively larger team within one institution \citep{dong2018collaboration, jones2008multi}. Here, we find that China's team composition is close to that of other countries when the team size is small, demonstrating a similar extent of institute diversity (Fig. 3a and S5). However, different from other countries, China's institute diversity increases much slower as the team size increases. A significant fraction of big teams remain to be formed by members from the same institute (Fig. 3b, c and S5). For example, for all China's papers by 6 authors in 2017, nearly 50\% of them are done in the same institute, which is 14 percentage points higher than that of the United States. A similar conclusion also holds when we use the fraction of papers done by no more than 2 institutes. It is encouraging to notice the trend that teams tend to be more diverse as time goes. The one-institute papers take a fewer percentage of total publications now than in the past (Fig. S6). However, the rate of change is low, suggesting that the institute diversity for China, an important factor for innovative works, will not improve very much in the near future.

\begin{figure}[htp]
\centering
\includegraphics[width=1.00\textwidth]{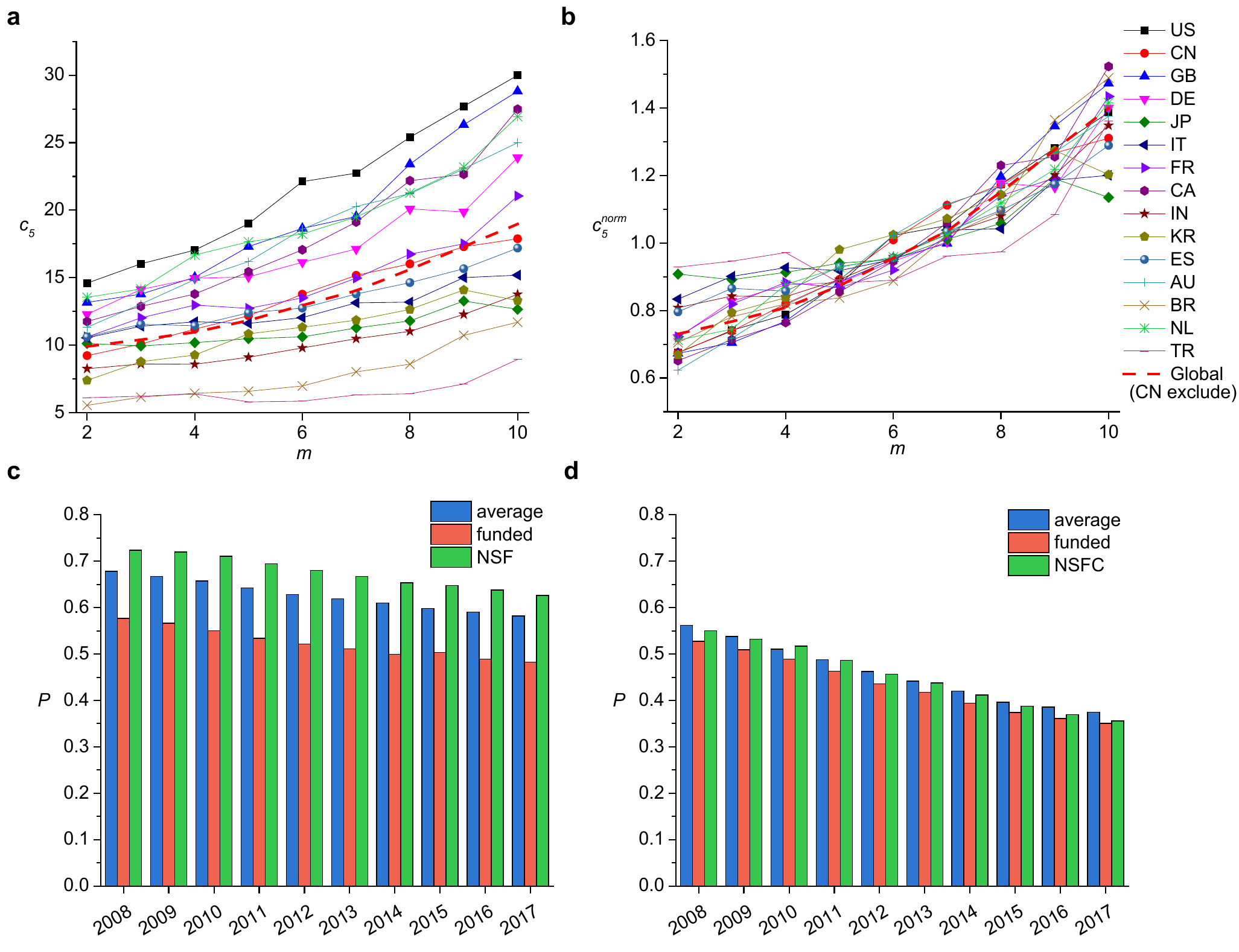}
\caption{{\bf (a)} The total number of citations a paper receives within 5 years of its publication $c_5$ is positively correlated with the team size $m$ in every country. The statistics are based on papers published in 2011. {\bf (b)} When re-scaling the number of citation of $c_5$ by the average value of a country as $c_5^{norm} = c_5 / \langle c_5 \rangle$, different curves in (a) almost collapse to a single curve showing similar increasing trend with team size. {\bf (c)} The fraction of small team output ($m \le 4$) among all papers, funded papers, and papers supported by NSF from United States. While funding agencies in general prefer big teams, NSF supports more small team works than the average.  {\bf (d)} The fraction of small team output ($m \le 4$) among all papers, funded papers, and papers supported by NSFC from China. In 2011, NSFC supports small team works slightly more than the national average. In 2010, the difference between NSFC and the national average is not statistically significant. In most of the time, NSFC supports fewer small team works than the average.
}
\label{fig:fig4}
\end{figure}

So far, we have demonstrated the aspects that China differ from other counties in research teams. What remains unclear are the factor that gives rise to the differences observed. Given confounding factors in team assembling in different countries, identifying these factors is out of this paper's scope. Nevertheless, we perform some preliminary analyses by proposing and testing two hypotheses that seem capable of explaining the observations.\\
H1: Papers by big teams have a higher capability in receiving more citations in China than in other countries.\\
H2: Big teams are more preferred by funding agencies in China than in other countries.\\

Note that papers by larger teams on average receive more citations than those by smaller teams \citep{klug2016understanding, wu2019large, wuchty2007increasing}. The argument for H1 is that the citation boost is more considerable in China, which consequently provides incentives to build large teams. We count the total number of citations a paper receives within 5 years of its publication $c_5$, and find that $c_5$ overall positively correlates with the team size $m$ (Fig. 4a). Papers from different countries receive different levels of citations. However, after re-scaling these curves by the national average of $c_5$, they almost collapse to a single curve (Fig. 4b and Fig. S7). The trend that papers by bigger teams receive higher citations is not different, at least not more extreme, in China than in other countries. The same conclusion also holds when we use a shorter time window to count citations (Fig. S8). Hence we conclude that H1 is not supported by the data.

We test H2 by extracting the grant information of each paper. Over 80\% of papers from China contains the grant information, much higher than other countries (Fig. S9). It implies that Chinese scientists are more obligated to acknowledge the funding agencies, or simply that only teams capable of securing research grants can efficiently conduct scientific research \citep{wang2019early, yang2015matthew, wang2012science}. Either of these explanations suggests the significant impact of funding agencies on scientific research in China. As intuitively expected, the percentage of papers with grants increases with team size in almost every country (Fig. S9). But once again, the increase is not sharper in China than in other countries (Fig. S9). Nor could we find any difference in the number of grants a paper is supported by (Fig. S10).

Indeed, given different sources of funding in different nations, different policies and aims of different funding agencies, and potential flaws in the records that may affect the observation \citep{alvarez2017funding, paul2016characterization, tang2017funding,azoulay2011incentives}, it would be less meaningful to test H2 by comparing all grants and papers from all countries. For this reason, we then consider only two grants: the National Natural Science Foundation of China (NSFC) and the National Science Foundation of the United States (NSF). It is believed that China learned from NSF to initiate and organize NSFC. They two have a very similar amount of budget (especially after taking purchasing power into consideration), scope and aim. In addition, both of them are one of the major national funding resources for fundamental research \citep{huang2016does, wang2012science}. All these features make NSFC and NSF two comparable examples. For each of China and the United States, we collect three sets of papers: all papers published in a given year, papers supported by grants in that year, and papers mainly supported by NSFC or NSF in that year (see Data and Methods for details). Compared with the national average, the small team output is less in papers with grants (Fig. 4c, d, S11 and S12), in line with our previous finding that works by larger teams have a higher probability of being sponsored. However, within papers supported by NSF, the percentage of output from small teams is higher than average (Fig. 4c, S11 and S12). On the contrary, the fraction of small team output is usually less than average in papers supported by NSFC (Fig. 4d, S11 and S12). In other words, using the national average as the baseline, NSF supports more small team works than NSFC does. Given the similarities between the two funding agencies, this observation supports H2.

It may be argued that NSFC and NSF are not comparable because China does not have an independent funding agency like the National Institutes of Health (NIH) that mainly focuses on biomedical research. Therefore, NSFC supports more works in medicine that rely mainly on corporations by big teams. Consequently, the percentage of small team output is dragged down. Statistically, the argument stands. The fraction of supported works in biology is roughly the same for NSFC and NSF, where 12\% of NSFC supported works and 13.5\% of NSF supported works are in biology. But there is a non-negligible difference in the field of medicine: 9.6\% of NSFC supported works are in medicine while this value is only 2.4\% for NSF. Such difference by itself is related to intriguing questions in research management and policy, as it is unclear if combining application-oriented research like medicine with the basic research, like what NSFC does, would enhance the efficiency. Nevertheless, in terms of data analyses, we can do a treatment in the data by excluding NSFC and NSF supported papers in the field of medicine. After this modification, papers supported by NSFC that are carried out by small teams are slightly more than the national average (Fig. S14 and S15). The extent that NSFC is over the national average, however, is still smaller than that of NSF does. Hence, even after excluding papers in medicine, NSF supports more small team works than NSFC does, supporting our conclusion above.

\section{Conclusion}
To summarize, we analyze over 26 million papers on Web of Science published from 2000 to 2017, which is one of the most extensive analyses in terms of papers covered. We find that China's research output is more dominated by big teams than the rest of the world. The fraction of papers by small teams in China is not only much lower than the global average, ranking the last among the top 15 countries of scientific publications in 2017, but also has undergone a much stepper decrease since 2000. More importantly, as teams in China shift from small to large size, the team diversity that is essential for innovative works does not follow the same increase as that in other countries. A high percentage of works are carried out within one or two institutes. All of these observations indicate that China is very different from other countries in the composition of big and small teams in scientific research. If referring to the global average or country like the United States, China is way apart from the balance point.

Given the importance of the problem, we also make some preliminary attempts to understand factors that explain the different small/big team composition in China. The first hypothesis we test is that China's big teams have a more considerable advantage to gain citations than that of other countries. Hence there are more incentives to build a large team. Indeed, works by larger teams on average receive more citations than those by small teams. However, the citation boost is roughly the same in every country after taking the national average citation into consideration, implying that citation alone can not explain the difference. We then turn to check if large teams are more preferred by funding agencies. More than 80\% of papers from China acknowledge research grants, which is the highest among the 15 countries analyzed. It clearly indicates the significant influence of funding agencies on China's scientific research. While works by large teams are more apt to be supported by grants, China does not demonstrate any different patterns on this matter, following the same trend as other countries. Yet, when we separately compare the works supported by NSFC and NSF, we find that NSFC supports fewer small team works than NSF does. This gives some clues supporting the hypothesis that preferences by the funding agencies may be associated with the imbalance of small and big teams in China.

The concern on the balance between small and big teams is a relatively new topic, which was rarely studied in the past. Nevertheless, if we admit that small and large teams play different yet equally essentials roles in scientific research, we need to consider the imbalance seriously. Our analyses based on the large volume of publication data bring evidence suggesting that China may need more small team output. If the drop of small teams persists, China may become less competitive in delivering disruptive research outcome and expanding the frontier of the field. One day, science community in China may not have enough new questions for its big teams to further develop and exploit.
The factor we spot that is associated with this imbalance further sheds light on this issue. Giving multiple confounding factors that may influence the organization of teams, we admit that our finding is preliminary. For example, one fundamental assumption of this study is that the patterns observed in big and small teams are universal. It is, however, reasonable to question this assumption. Indeed, if China's big teams are as capable as small teams in performing disruptive research, or if team diversity are not correlated with the impact of the work in China, the results reported in this paper would raise little worry. Currently we have some preliminary results confirming that patterns in big and small teams are universal, providing the basis of the research. But the nationality and universality in the science of science study is an interesting future direction. 
Some observations in this paper can be explained by the fact that big teams in China are more productive than those in other countries. However, given the fluidness in team assembly \citep{wang2015scientific, milojevic2014principles, abramo2017relationship}, testing this hypothesis is challenging, which requires a better author name disambiguation algorithm and other techniques to extract the core of the team in the scientific collaboration network \citep{wang2020measuring, yu2019academic}. The lack of team diversity and the large average team size in China can also be associated with the honorary authorship, where scholars not directly contributing to the work are added to the author list \citep{biagioli2018academic}. Although there is no evidence that such misconduct is more severe in China \citep{tang2019five}, the effects of honorary authorship on the team size may be worth further investigations. The collectivist culture in Asia may also encourage the formation of big teams. Both Japan and Korea have a relatively small percentage of small team output. Exploring these factors may not only provide useful insights to the research community in China, but also advance our quantitative understanding of science \citep{fortunato2018science,azoulay2018toward}.

\newpage

	\bibliography{team}
	
	\section{Supportive Information}
	See the file of supplementary information for additional statistics and figures.

	\acknowledgments
	We thank Prof. Barabasi at CCNR for giving access to the WOS data. The work is supported by the National Natural Science Foundation of China (No. 61603309).

	\authorcontributions 
	T.J., J.H., and F.X. designed the research, J.H. and T.J. did the data parsing and cleaning, T.J., L.L. and F.Y analyzed the data, collected the statistics and reviewed related literature. T.J. and L.L. prepared the initial draft of the manuscript. All authors contributed comments on the results and revisions to the final version.
	\clearpage

\hypersetup{CJKbookmarks,%
	bookmarksnumbered,%
	colorlinks,%
	linkcolor=black,%
	citecolor=black,%
	plainpages=false,%
	pdfstartview=FitH}

\renewcommand{\figurename}{{\bf Supplementary Figure}}
\renewcommand{\thefigure}{{\bf S\arabic{figure}}}
\renewcommand{\thesection}{S\arabic{section}}
\renewcommand{\thesubsection}{S\arabic{section}.\arabic{subsection}}
\renewcommand{\tablename}{{\bf Supplementary Table}}
\renewcommand{\thetable}{{\bf S\arabic{table}}}
\renewcommand{\theequation}{S\arabic{equation}}
\renewcommand\thefigure{S\arabic{figure}} 
\def\msec#1{\bigskip\textbf{#1}}
\def\note#1{{\small\color{red}\textbf{[[#1]]}}}
\setlength{\LTcapwidth}{\textwidth}

	\section{Supplementary Information}
	\begin{figure}[h]
		\begin{center}
			\setcounter{figure}{0}
			\resizebox{13cm}{!}{\includegraphics{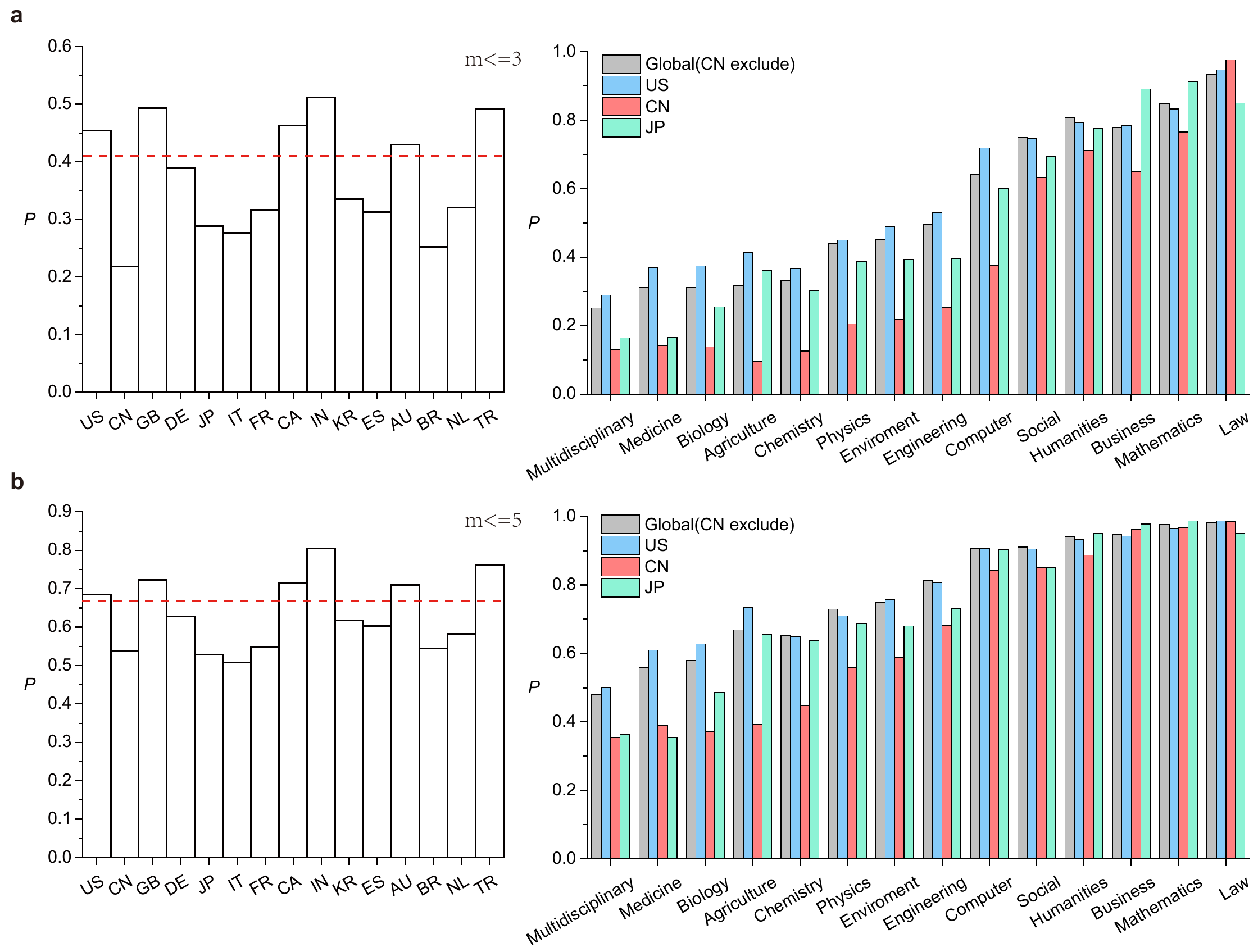}}
			\caption{{\bf (a)} The fraction of papers in 2017 produced by teams with size no more than 3 ($P(m \le 3)$) in different countries and in different fields. {\bf (b)} The fraction of papers in 2017 produced by teams with size no more than 5 ($P(m \le 5)$) in different countries and in different fields. }
			\label{fig:S1}
		\end{center}
	\end{figure}\noindent 
	
	\clearpage
	
	\begin{figure}[ht]
		\begin{center}
			\resizebox{12cm}{!}{\includegraphics{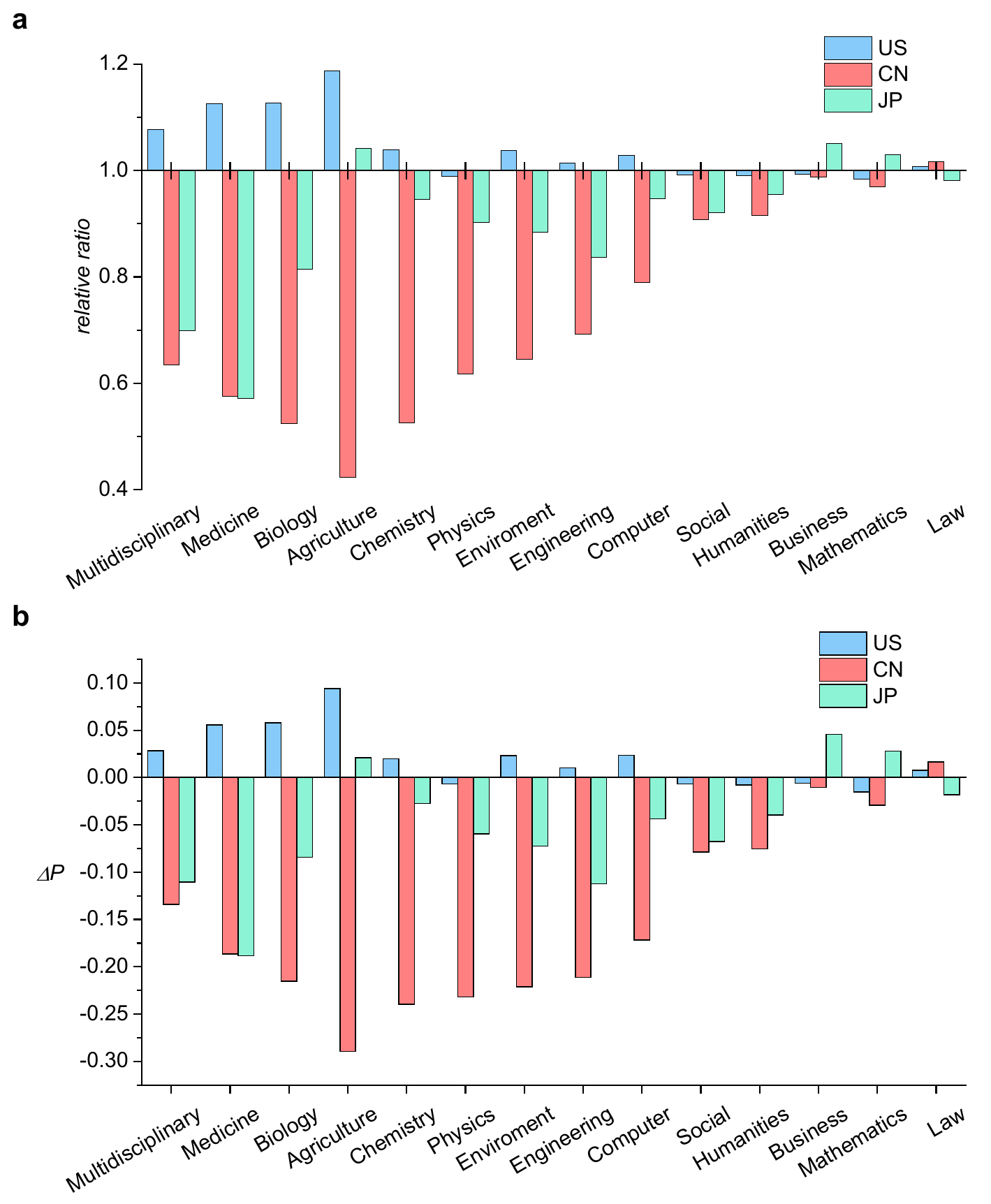}}
			\caption{{\bf (a)} The relative ratio of papers (setting the global average as y=1) in 2017 produced by teams with size no more than 4 ($P(m \le 4)$) in different fields. {\bf (b)} The absolute ratio of papers (setting the global average as y=0) in 2017 produced by teams with size no more than 4 ($P(m \le 4)$) in different fields.}
			\label{fig:S2}
			
		\end{center}
	\end{figure}\noindent 
	
	\clearpage
	
	\begin{figure}[ht]
		\begin{center}
			\resizebox{13cm}{!}{\includegraphics{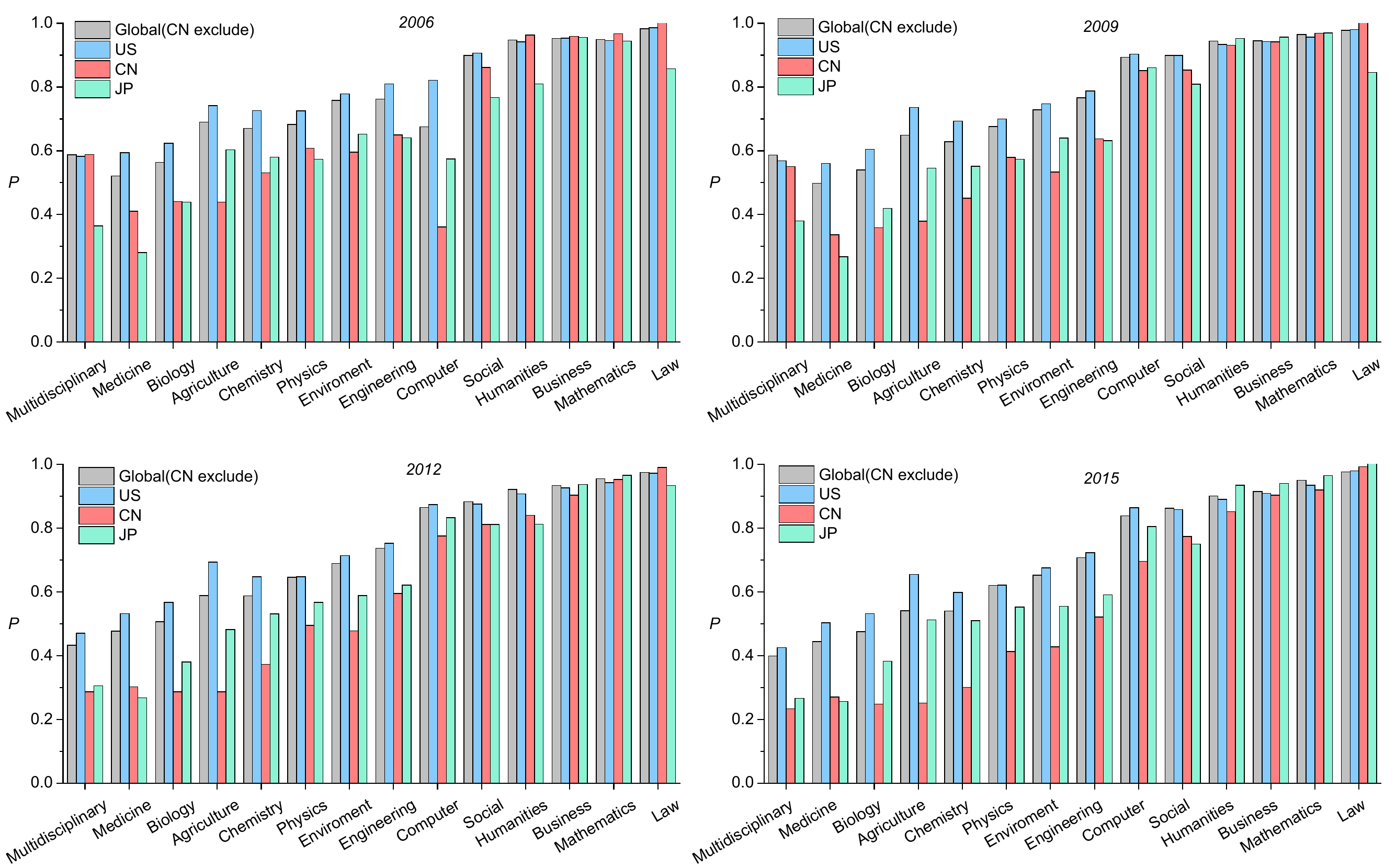}}
			\caption{The fraction of papers produced by teams with size no more than 4 ($P(m \le 4)$) in different years and in different fields. }
			\label{fig:S3}
			
		\end{center}
	\end{figure}\noindent 
	
	\clearpage
	\begin{figure}[ht]
		\begin{center}
			\resizebox{13cm}{!}{\includegraphics{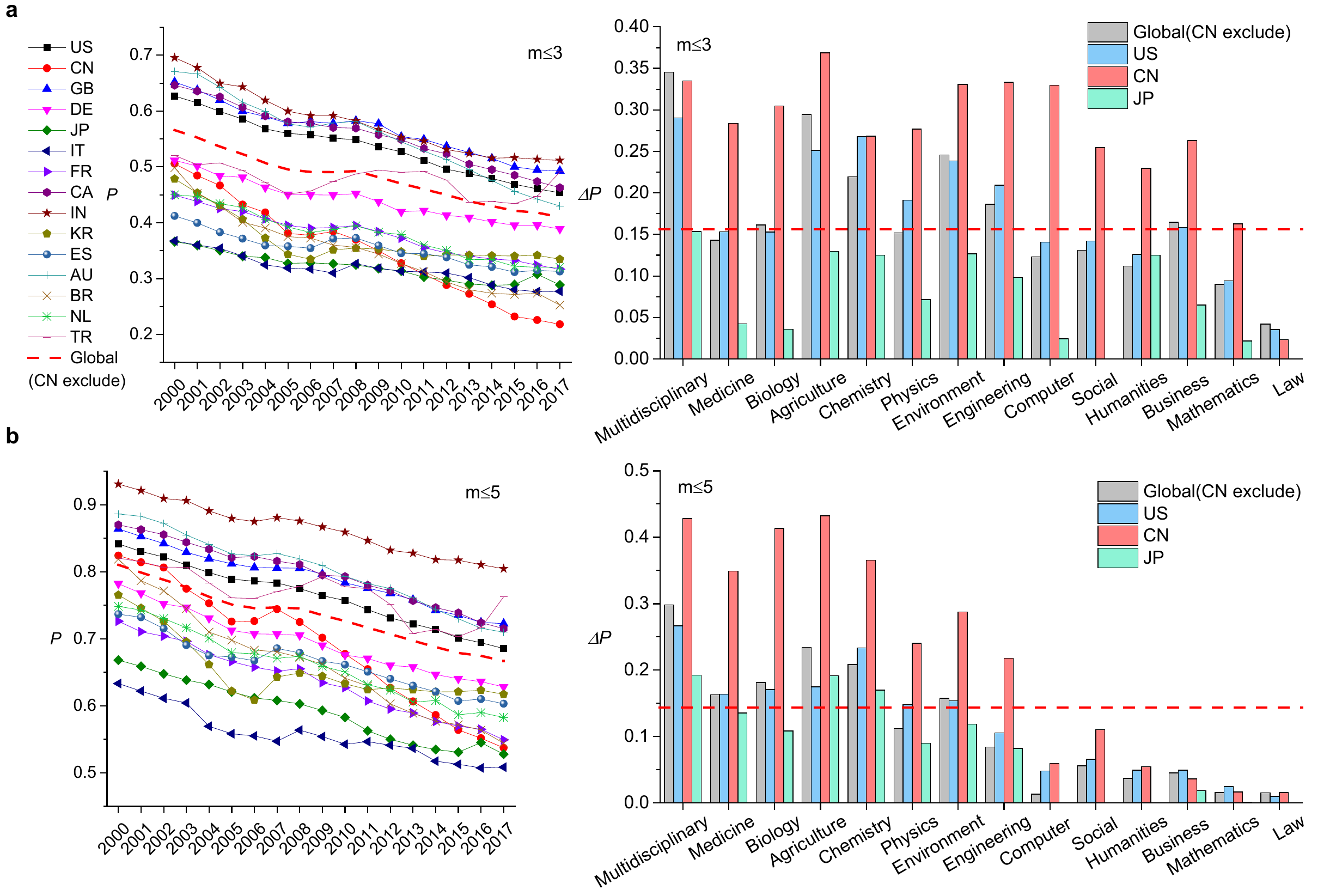}}
			\caption{{\bf (a)} The time evolution of $P(m \le 3)$ in different countries and the drop of $P(m \le 3)$ from year 2000 to 2017 in different fields. {\bf (b)} The time evolution of $P(m \le 5)$ in different countries and the drop of $P(m \le 5)$ from year 2000 to 2017 in different fields.  }
			\label{fig:S4}
			
		\end{center}
	\end{figure}\noindent 
	
	\clearpage
	
	\begin{figure}[ht]
		\begin{center}
			\resizebox{13cm}{!}{\includegraphics{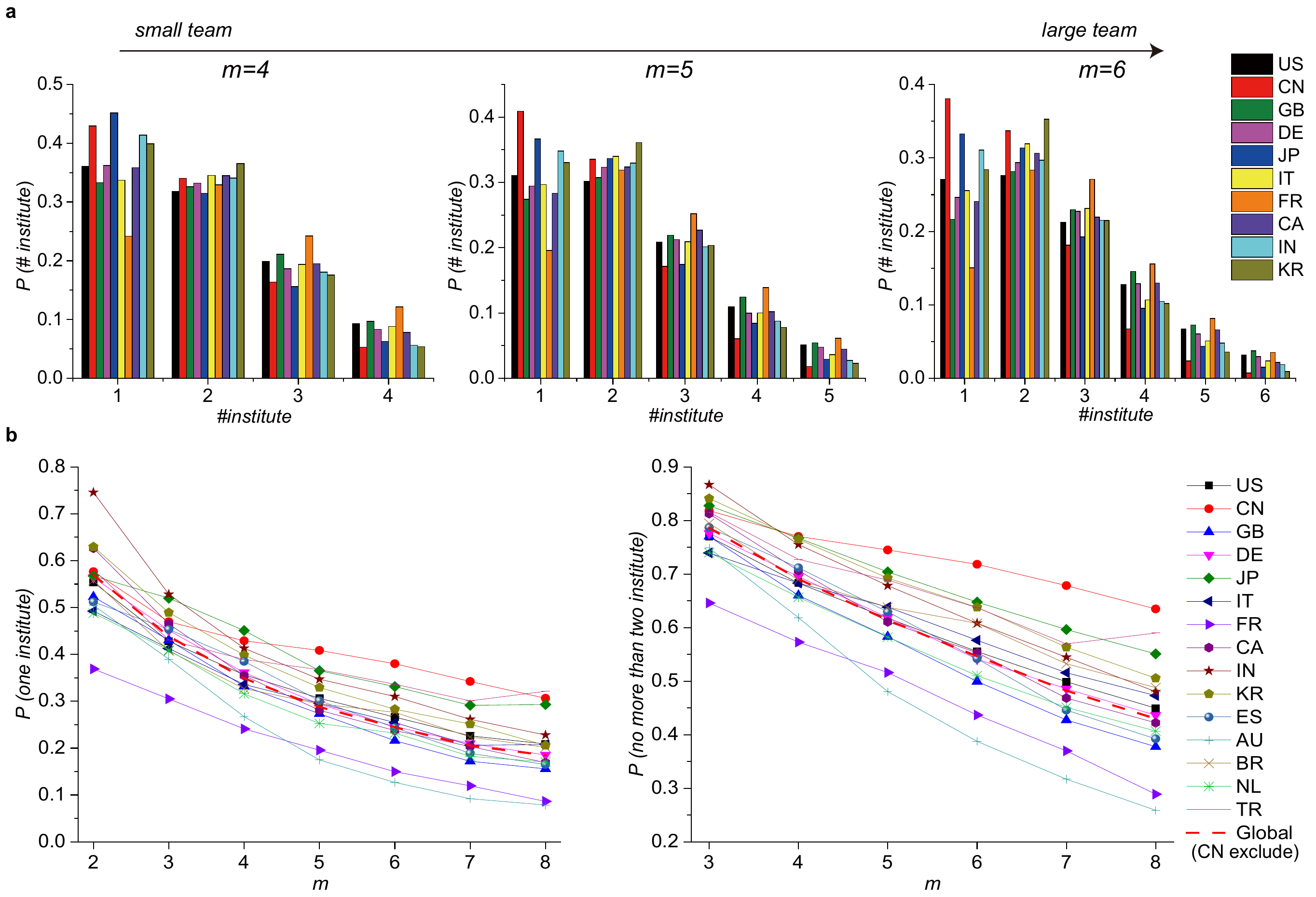}}
			\caption{The results based on the \textbf{paper’s afﬁliation}. {\bf (a)} The distribution of the number of distinct institutes in papers published in 2017 with team size $m = $ 4, 5  and 6 (from left to right). {\bf (b)} The fraction of papers in 2017 done by one institute, given the team size $m$. More percentage of paper output is from a single institute in China than in other countries. {\bf (c)} Similar to (b). The fraction of papers in 2017 involving no more than two institutes. }
			\label{fig:S5}
		\end{center}
	\end{figure}\noindent 
	
	\clearpage
	
	\begin{figure}[ht]
		\begin{center}
			\resizebox{13cm}{!}{\includegraphics{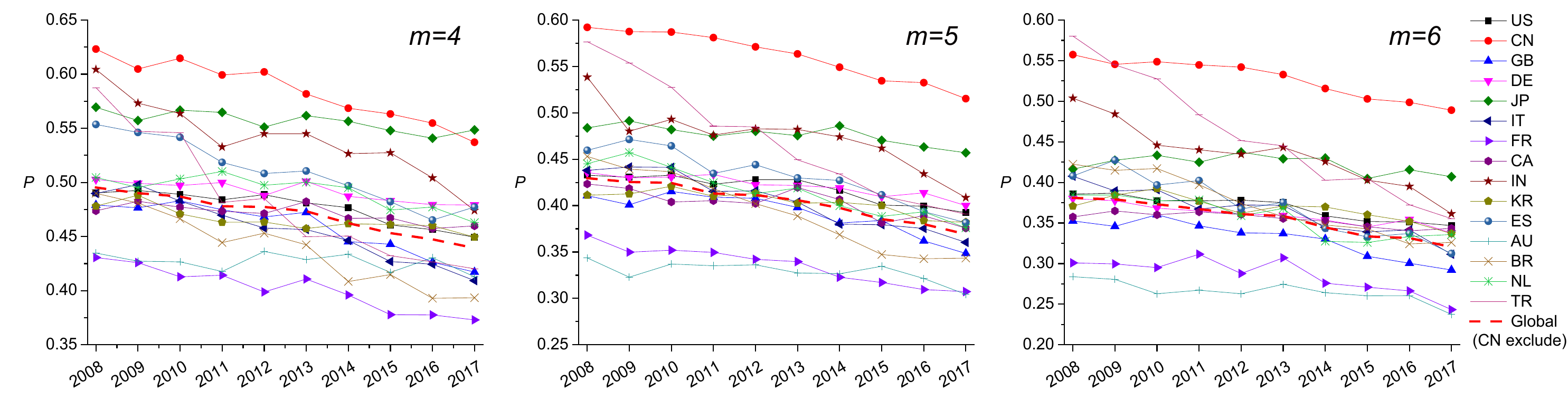}}
			\caption{The fraction of papers done by one institute from 2008 to 2017, with the team size $m=$ 4, 5, 6(from left to right). }
			\label{fig:S6}
		\end{center}
	\end{figure}\noindent 
	
	\clearpage
	
	\begin{figure}[ht]
		\begin{center}
			\resizebox{13cm}{!}{\includegraphics{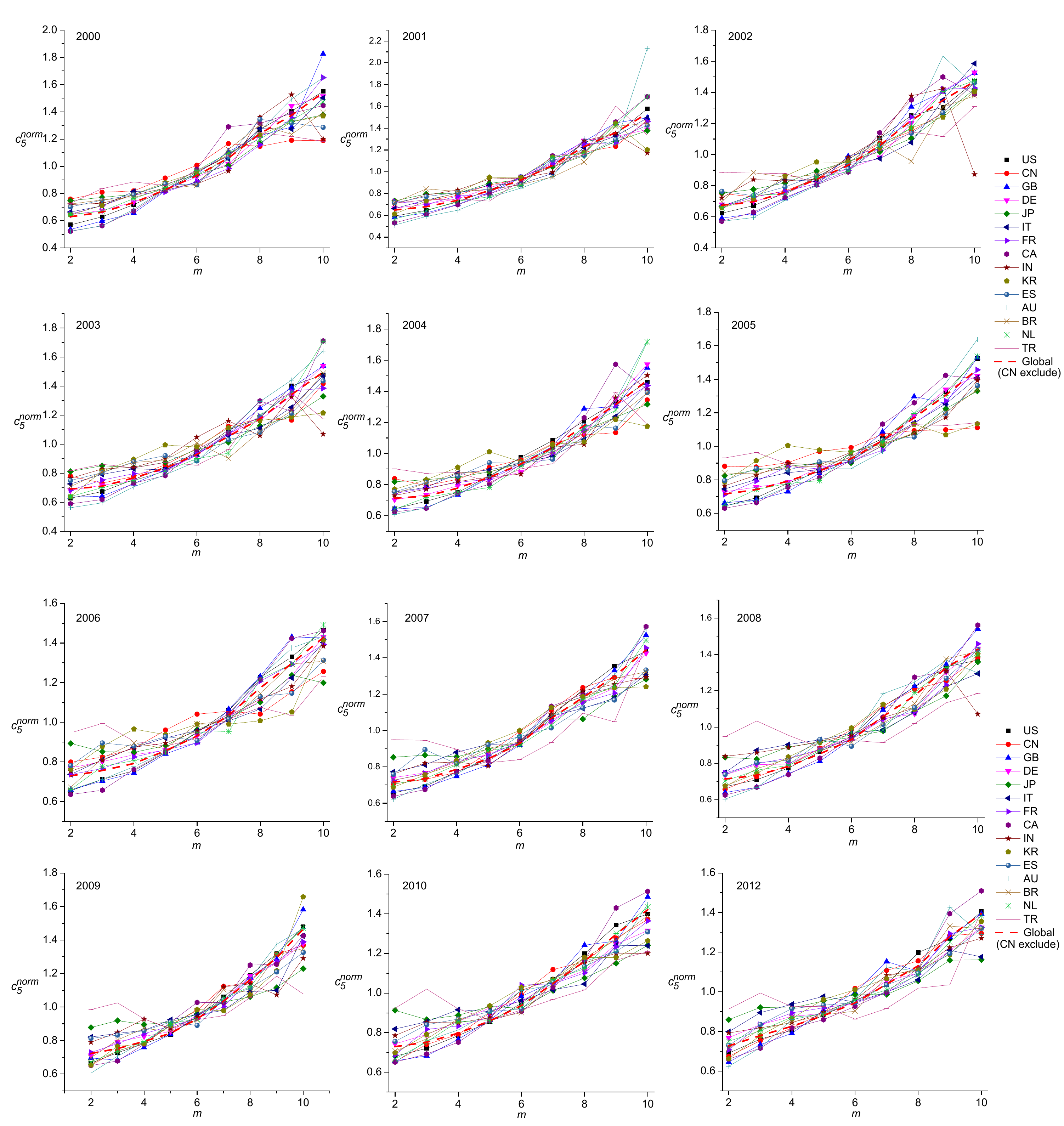}}
			\caption{Re-scaling the number of citation of $c_5$ by the average value of a country as $c_5^{norm} = c_5 / \langle c_5 \rangle$ in different years. }
			\label{fig:S7}
		\end{center}
	\end{figure}\noindent 
	
	\clearpage
	
	\begin{figure}[ht]
		\begin{center}
			\resizebox{13cm}{!}{\includegraphics{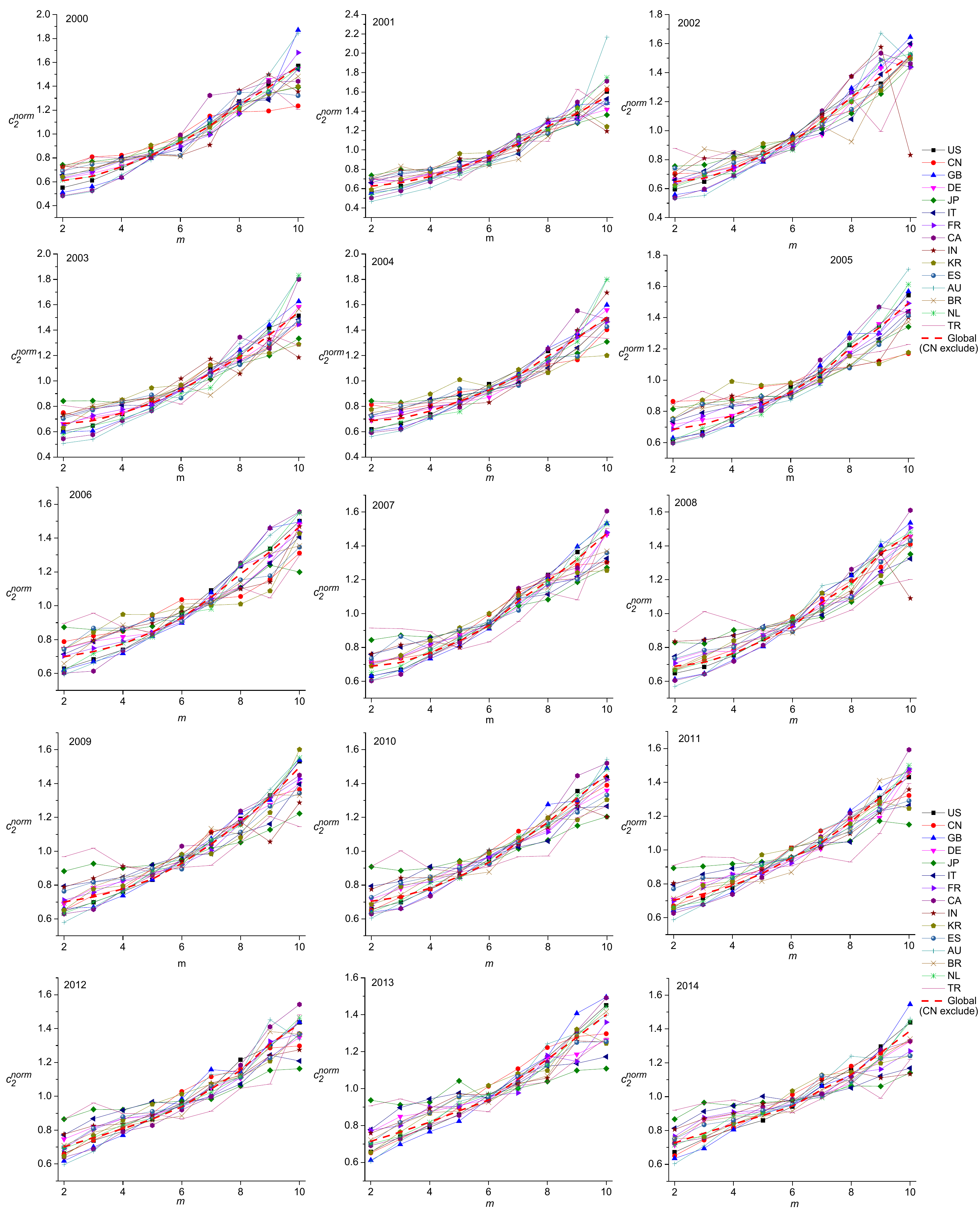}}
			\caption{Re-scaling the number of citation of $c_2$ by the average value of a country as $c_2^{norm} = c_2 / \langle c_2 \rangle$ in different years. }
			\label{fig:S8}
		\end{center}
	\end{figure}\noindent 
	
	\clearpage
	
	\begin{figure}[ht]
		\begin{center}
			\resizebox{13cm}{!}{\includegraphics{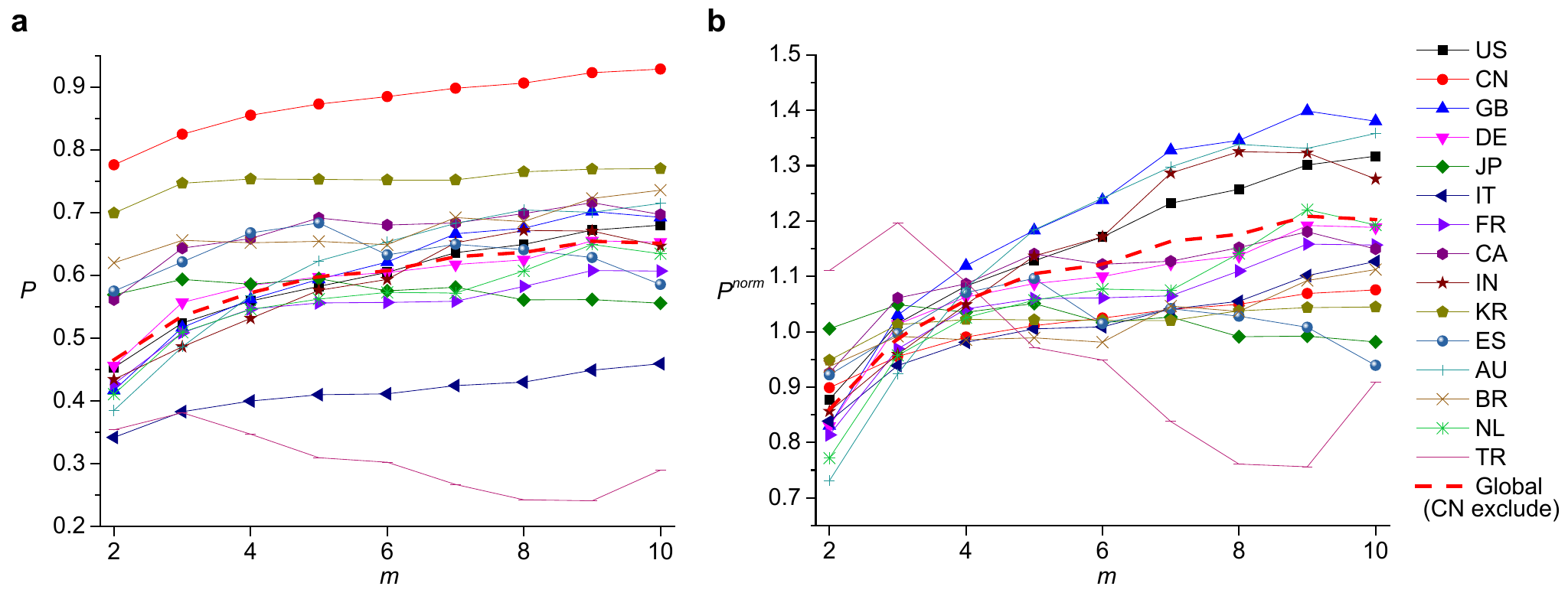}}
			\caption{{\bf (a)} The relationship between the fraction of papers acknowledge funding support and the team size m. The percentage of funded papers are highest in China.  {\bf (b)} Curves in (a) is re-scaled by the average value of a country. Large teams in China do not have a higher than global average trend to have works supported by research grants. }
			\label{fig:S9}
		\end{center}
	\end{figure}\noindent 
	
	\clearpage
	
	\begin{figure}[ht]
		\begin{center}
			\resizebox{13cm}{!}{\includegraphics{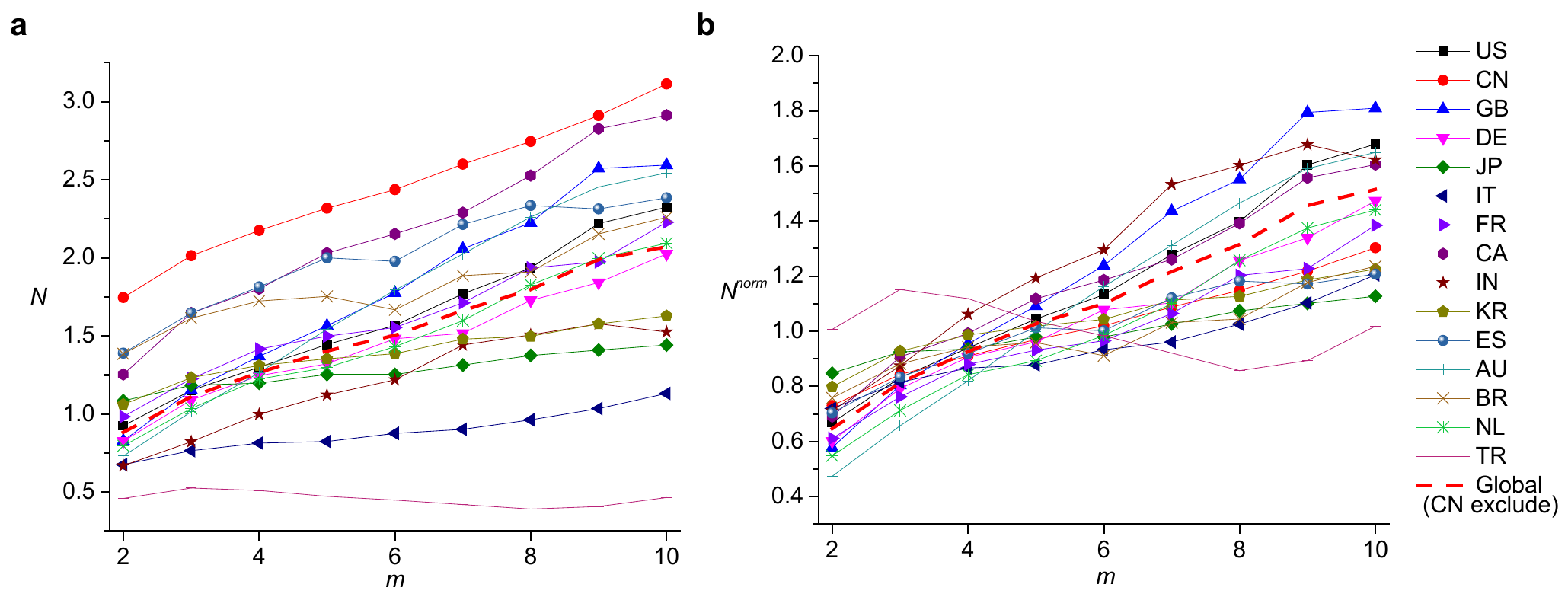}}
			\caption{{\bf (a)} The relationship between the fraction of papers acknowledge funding support and the team size $m$. The number of grants a paper of funded papers are highest in China. {\bf (b)} Curves in (a) is re-scaled by the average value of a country. Large teams in China do not have a higher than global average trend to have works supported by research grants. }
			\label{fig:S10}
		\end{center}
	\end{figure}\noindent 
	
	\clearpage
	
	\begin{figure}[ht]
		\begin{center}
			\resizebox{13cm}{!}{\includegraphics{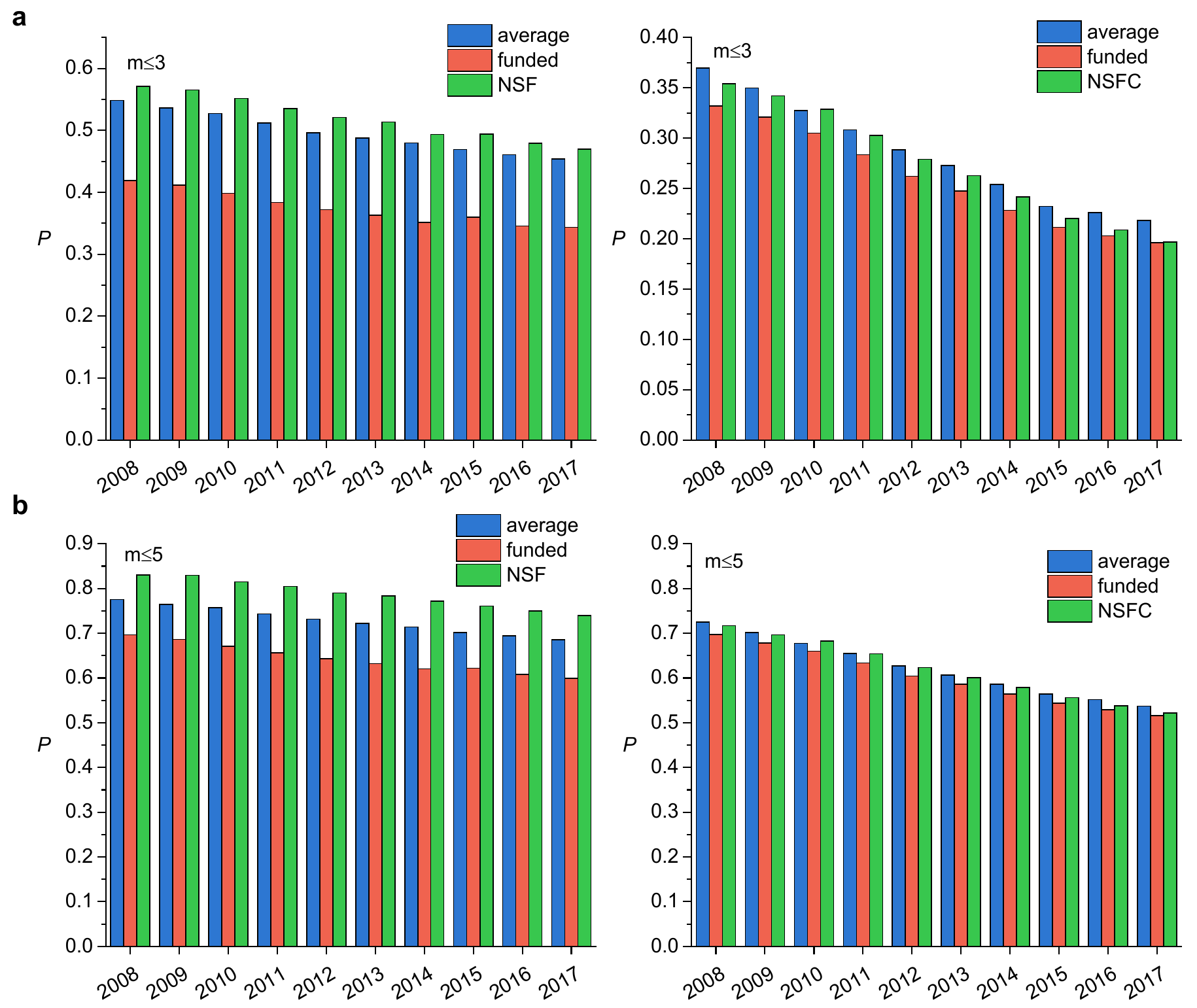}}
			\caption{ {\bf (a)} The fraction of small team output ($m \le 3$) among all papers, funded papers, and papers supported by NSF from United States and NSFC from China. {\bf (b)} The fraction of small team output ($m \le 5$) among all papers, funded papers, and papers supported by NSF from United States and NSFC from China. }
			\label{fig:S11}
		\end{center}
	\end{figure}\noindent 
	
	\clearpage
	
	\begin{figure}[ht]
		\begin{center}
			\resizebox{13cm}{!}{\includegraphics{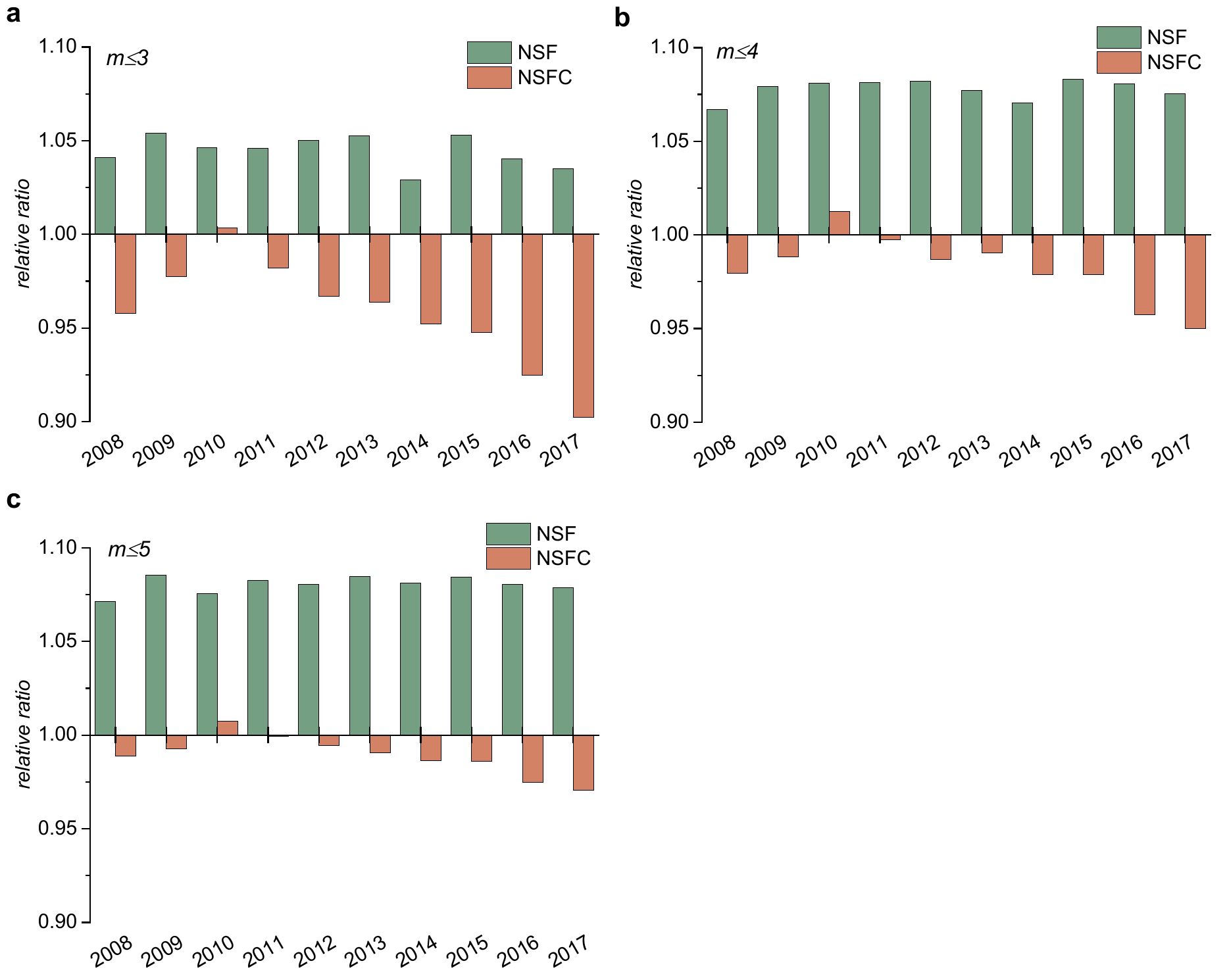}}
			\caption{ {\bf (a)} The relative ratio (setting all papers as y=1) of small team output ($m \le 3$) supported by NSF from United States and NSFC from China. {\bf (b)} The relative ratio (setting all papers as y=1) of small team output ($m \le 4$) supported by NSF from United States and NSFC from China.{\bf (c)} The relative ratio (setting all papers as y=1) of small team output ($m \le 5$) supported by NSF from United States and NSFC from China.}
			\label{fig:S12}
		\end{center}
	\end{figure}\noindent 
	
	\clearpage
	
	\begin{figure}[ht]
		\begin{center}
			\resizebox{13cm}{!}{\includegraphics{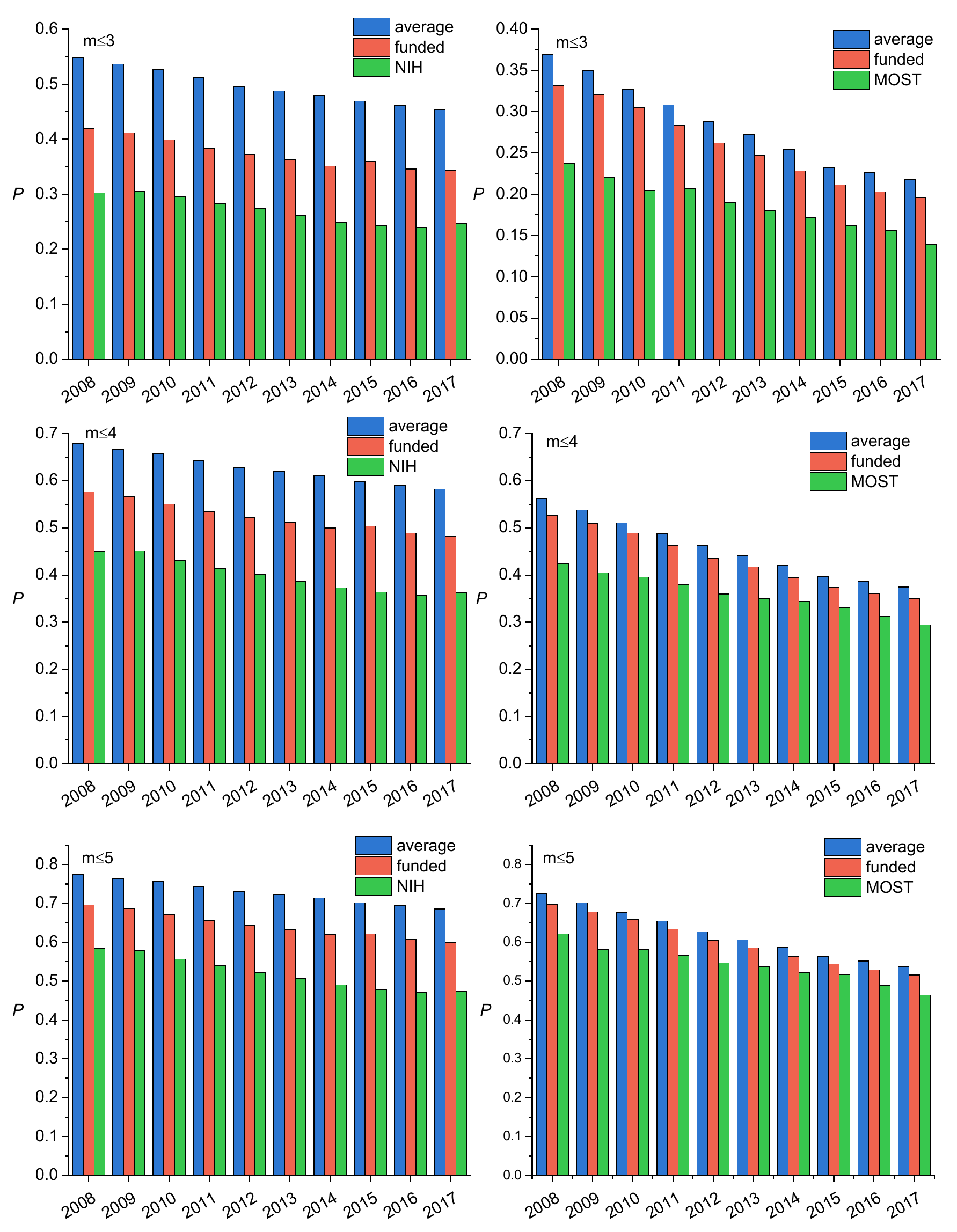}}
			\caption{The fraction of small team output ($m \le 3$, $m \le 4$, $m \le 5$) among all papers, funded papers, and papers  supported by NIH from United States and MOST from China. }
			\label{fig:S13}
		\end{center}
	\end{figure}\noindent 
	
	\clearpage
	
	\begin{figure}[ht]
		\begin{center}
			\resizebox{13cm}{!}{\includegraphics{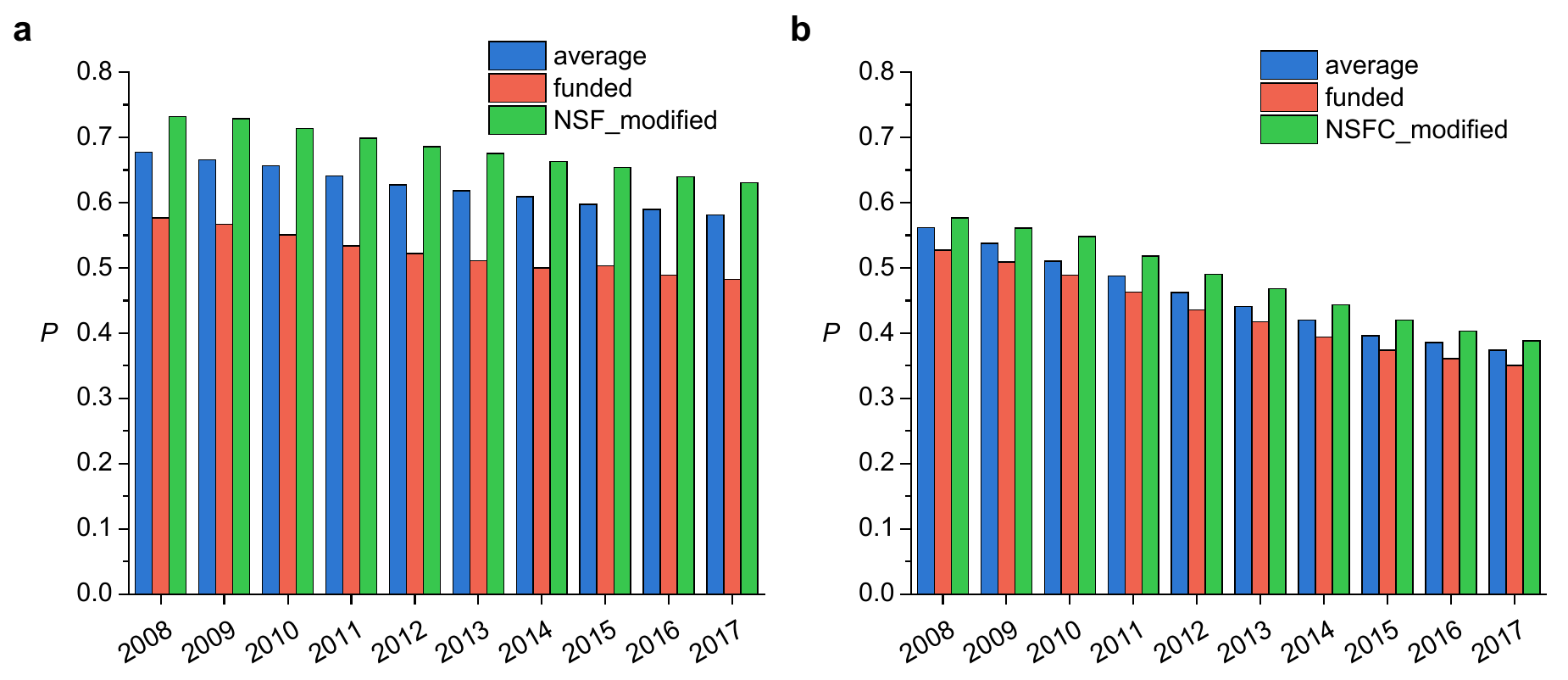}}
			\caption{ {\bf (a)} The fraction of small team output ($m \le 4$) among all papers, funded papers, and papers \textbf{exclude Medicine} supported by NSF from United States. {\bf (b)} The fraction of small team output ($m \le 4$) among all papers, funded papers, and papers \textbf{exclude Medicine} supported by NSFC from China. }
			\label{fig:S14}
		\end{center}
	\end{figure}\noindent 
	
	\clearpage
	
	\begin{figure}[ht]
		\begin{center}
			\resizebox{13cm}{!}{\includegraphics{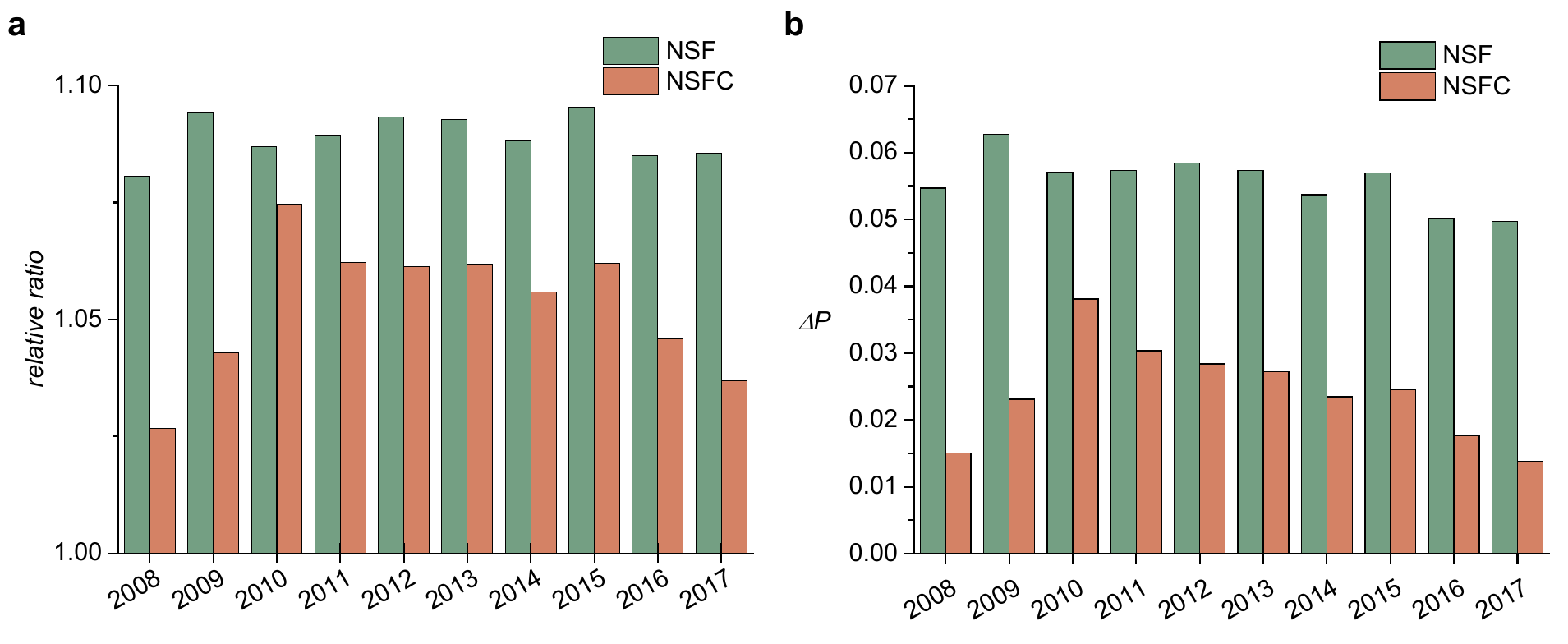}}
			\caption{{\bf (a)} The relative ratio (setting all papers as y=1) of small team output ($m \le 4$) \textbf{exclude Medicine} supported by NSF from United States and NSFC from China. {\bf (b)} The absolute ratio (setting all papers as y=0) of small team output ($m \le 4$) \textbf{exclude Medicine} supported by NSF from United States and NSFC from China. }
			\label{fig:S15}
		\end{center}
	\end{figure}\noindent 
	
	\clearpage
	
	{\bf Supplementary Note 1}
	
	{\bf Results based on publications by authors from the same country.}
	The fraction of papers in 2017 produced by teams with size no more than 4 (P($m \le 4$)) in China ranks the last among the top 15 countries of scientiﬁc papers (Fig.S16a). We further analyze the P(($m \le 4$)) in different research ﬁelds (Fig.S16b). We find that the result is similar to Figure.1. Then we ﬁnd in our analyses that the percentage of papers by small teams decreases over years. Nevertheless, the drop of China is much steeper. This result is similar to Figure.2 (Fig.S17).

	\begin{figure}[ht]
		\begin{center}
			\resizebox{13cm}{!}{\includegraphics{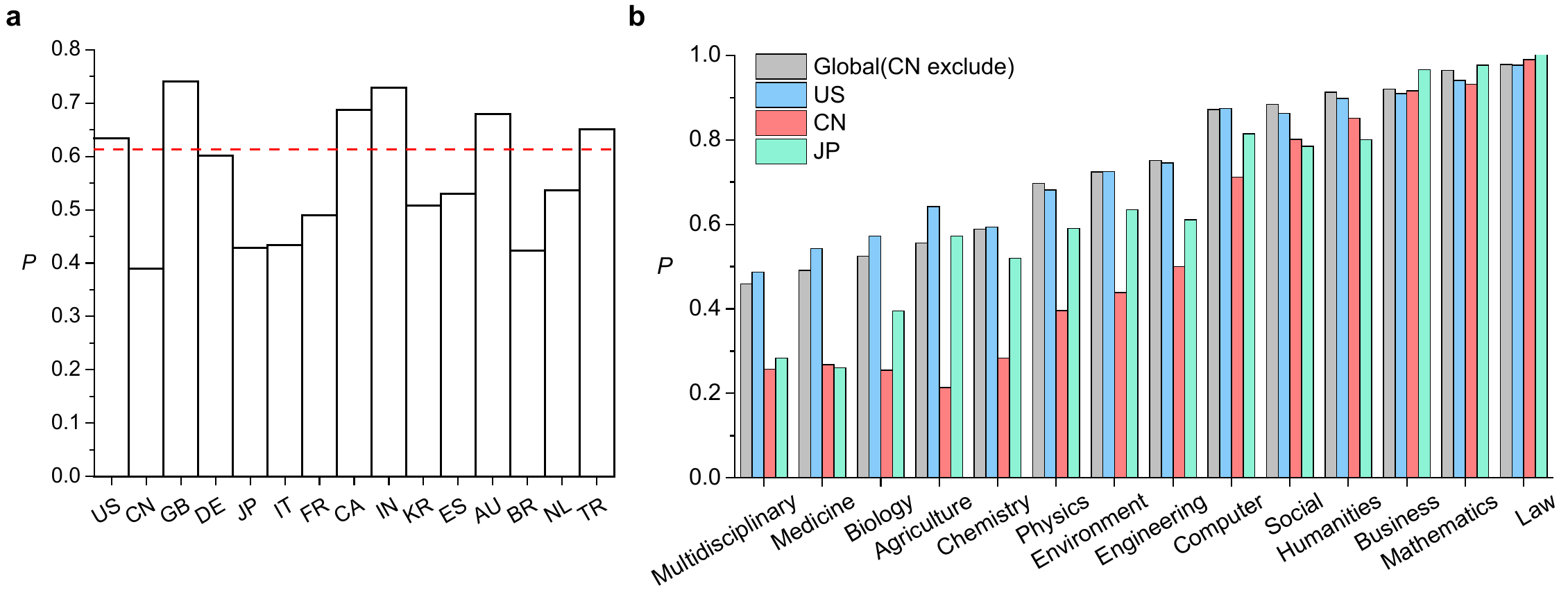}}
			\caption{{\bf (a)} The fraction of papers in 2017 produced by teams with size no more than 4 ($P(m \le 4)$) in different countries. The dashed line corresponds to the global average (in which China is excluded) {\bf (b)} $P(m \le 4)$ in different fields in the year 2017.}
		\end{center}
	\end{figure}\noindent 
	
	\clearpage
	
	\begin{figure}[ht]
		\begin{center}
			\resizebox{13cm}{!}{\includegraphics{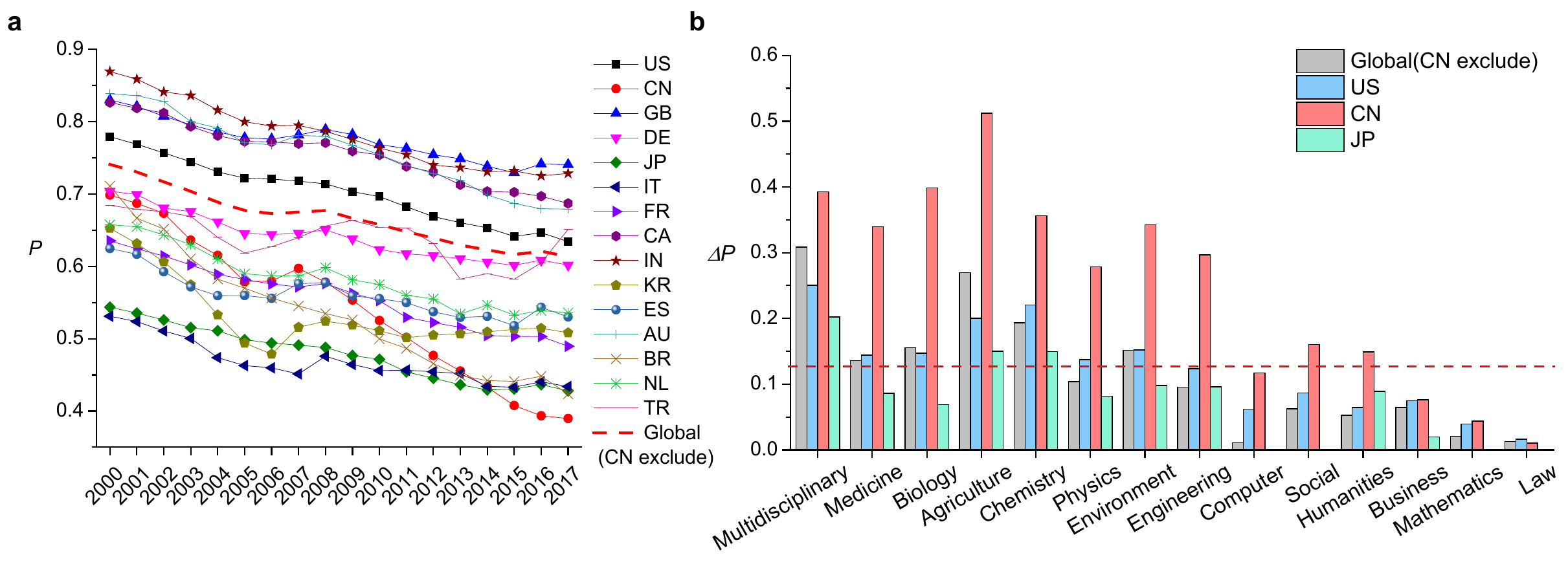}}
			\caption{{\bf (a)} The time evolution of $P(m \le 4)$ in different countries. The dashed line corresponds to the global average (in which China is excluded). {\bf (b)} The drop of $P(m \le 4)$ from year 2000 to 2017 in different fields. The dashed line corresponds to the drop of the global value in (a). }
		\end{center}
	\end{figure}\noindent 
	
	\clearpage
	
	{\bf Supplementary Note 2}
	
	{\bf China's decrease of the small team research over time significantly deviates from the global average.} The conclusion is based on two factors. First, the value of China and the global average are significantly different each year (see discussions in section "Statistical Test"). Therefore, any data points used for linear regression are distinct from each other.  Second, using the linear regression, the dropping slope of China is -0.01887, with a standard error $5.93286\times10^{-4}$. The dropping slop of the global average is -0.00869, with a standard error $3.04267\times10^{-4}$. The 95\% confidence interval of the dropping slope of China and global average do not overlap. Hence we conclude that China's deviation from the global average is statistically significant.

	\begin{figure}[ht]
		\begin{center}
			\resizebox{13cm}{!}{\includegraphics{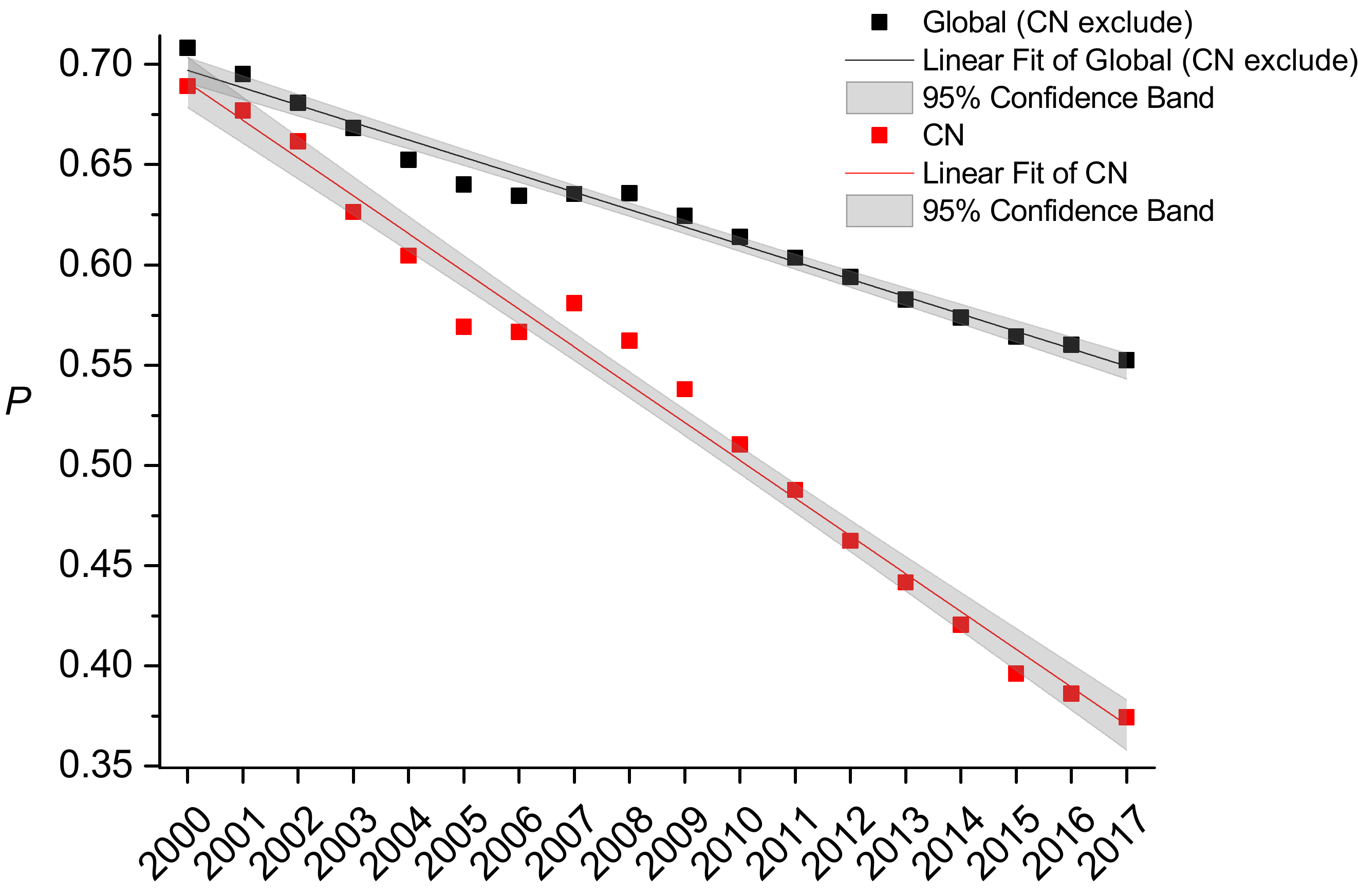}}
			\caption{ The linear regression on the drop of China and the drop of the global average. The two droppings do not overlap}
		\end{center}
	\end{figure}\noindent 
	
	\clearpage
	
	\begin{center}
		\begin{longtable}{|p{9cm}<{\centering}|p{3cm}<{\centering}|}
			\hline
			\leftline{\textbf{Names for grants from NSFC :}} & \textbf{No. of Occurrences}\\
			\hline%
			National Natural Science Foundation of China & 774174\\\hline%
			Natural Science Foundation of China & 61169\\\hline%
			NSFC & 59131\\\hline%
			National Science Foundation of China & 52912\\\hline%
			National Natural Science Foundation of China (NSFC) & 33614\\\hline%
			National Nature Science Foundation of China & 32077\\\hline%
			National Natural Science Foundation & 14868\\\hline%
			NSF of China & 10886\\\hline%
			NNSF of China & 5725\\\hline%
			National Natural Sciences Foundation of China & 5417\\\hline%
			National Natural Science Foundations of China & 5188\\\hline%
			Chinese National Natural Science Foundation & 4288\\\hline%
			National Science Foundation of China (NSFC) & 4119\\\hline%
			Natural Science Foundation of China (NSFC) & 3945\\\hline%
			National Natural Science Fund of China & 3498\\\hline%
			NNSFC & 2625\\\hline%
			Nature Science Foundation of China & 2621\\\hline%
			National Natural Scientific Foundation of China & 2526\\\hline%
			National Natural Science Funds of China & 1958\\\hline%
			National Natural Science Foundation of China (NNSFC) & 1895\\\hline%
			Natural Science Foundation & 1806\\\hline%
			NSFC of China & 1723\\\hline%
			National Natural Foundation of China & 1469\\\hline%
			Chinese Natural Science Foundation & 1467\\\hline%
			National Nature Science Foundation & 1459\\\hline%
			Chinese National Science Foundation & 1439\\\hline%
			National Nature Science Foundation of China (NSFC) & 1299\\\hline%
			National Natural Science foundation of China & 1248\\\hline%
			National Science and Technology Support Program of China & 1185\\\hline%
			China National Funds for Distinguished Young Scientists & 1183\\\hline%
			
			\leftline{\textbf{Names for grants from MOST :}} & \textbf{NO. of Occurrences} \\\hline%
			National Basic Research Program of China & 75523\\\hline%
			National Basic Research Program of China (973 Program) & 30924\\\hline%
			Ministry of Science and Technology of China & 20618\\\hline%
			National High Technology Research and Development Program of China & 16746\\\hline%
			National Key Basic Research Program of China & 9711\\\hline%
			National High Technology Research and Development Program of China (863 Program) & 8927\\\hline%
			111 Project & 8630\\\hline%
			National Key Research and Development Program of China & 7322\\\hline%
			National Key Technology RD Program & 5090\\\hline%
			National Basic Research Program & 4768\\\hline%
			National Key Technology R\&D Program of China & 3795\\\hline%
			National Science and Technology Major Project & 3239\\\hline%
			National Basic Research Program of China (973 program) & 3215\\\hline%
			973 Program of China & 3123\\\hline%
			Major State Basic Research Development Program of China (973 Program) & 3020\\\hline%
			Major State Basic Research Development Program of China & 3009\\\hline%
			973 program & 2674\\\hline%
			National Science and Technology Major Project of China & 2546\\\hline%
			National Key Basic Research Program of China (973 Program) & 2413\\\hline%
			Chinese Ministry of Science and Technology & 2395\\\hline%
			Ministry of Science and Technology of the People's Republic of China & 2389\\\hline%
			973 Project & 2200\\\hline%
			863 Program & 1964\\\hline%
			National Basic Research Program (973 Program) of China & 1704\\\hline%
			National Basic Research Program of China (973) & 1500\\\hline%
			National 973 Program of China & 1363\\\hline%
			National Basic Research Program (973) of China & 1347\\\hline%
			National 863 Program & 1267\\\hline%
			973 project & 1266\\\hline%
			National 973 Program & 1229\\\hline%
			National Basic Research Program (973 Program) & 1193\\\hline%
			National High-tech R\&D Program of China (863 Program) & 1187\\\hline%
			National 863 Program of China & 1178\\\hline%
			National 973 Project & 1071\\\hline%
			National Program on Key Basic Research Project (973 Program) & 1045\\\hline%
			
			\leftline{\textbf{Names for grants from NSF :}} & \textbf{NO. of Occurrences}\\
			\hline%
			National Science Foundation & 165984\\\hline%
			NSF & 119165\\\hline%
			National Science Foundation (NSF) & 17369\\\hline%
			U.S. National Science Foundation & 9569\\\hline%
			US National Science Foundation & 9191\\\hline%
			US National Science Foundation (NSF) & 1859\\\hline%
			US NSF & 1199\\\hline%
			NSF grant & 1192\\\hline%
			
			\leftline{\textbf{Names for grants from NIH :}} & \textbf{NO. of Occurrences}\\
			\hline%
			National Institutes of Health & 169423\\\hline%
			NIH & 161701\\\hline%
			National Institutes of Health (NIH) & 22227\\\hline%
			National Institute of Health & 12370\\\hline%
			US National Institutes of Health & 6838\\\hline%
			NIH/NHLBI & 1467\\\hline%
			National Institute of Health (NIH) & 1444\\\hline%
			NIH grant & 1194\\\hline%
			NIH/NCRR & 1044\\\hline%
			
			\caption{We extract the name of grant agency in each paper from China and the United States, filter out these appear fewer than 1000 times in the data, and manually identify names associated with National Natural Science Foundation of China (NSFC), Ministry of Science and Technology (MOST), National Institutes of Health and  National Science Foundation (NSF). These names and the number of occurrences are list in this Table. } \label{table1}	
			
		\end{longtable}

	\end{center}
	\clearpage
	
	\begin{center}
		\renewcommand\arraystretch{0.65}
		\begin{longtable}{|p{2cm}<{\centering}|p{2.5cm}<{\centering}|p{2.5cm}<{\centering}|p{4cm}<{\centering}|}
			\hline%
			Year & Funding Agency& No. of Occurrences & Percentage of Occurrences(\%)  \\
			\hline%
			\endfirsthead
			\endfoot
			\endlastfoot
			
			\multirow{4}{*}{2008} & NSFC & 16115 & 60.09 \\
			& MOST & 4021 & 14.99  \\\cline{2-4}
			& NSF & 10637 & 19.50 \\
			& NIH & 14578 & 26.73  \\\hline
			\multirow{4}{*}{2009} & NSFC & 49037 & 60.32\\
			& MOST & 13544 & 16.66 \\\cline{2-4}
			& NSF & 27225 & 19.45 \\
			& NIH & 36525 & 26.09 \\\hline%
			\multirow{4}{*}{2010} & NSFC & 59183 & 61.31 \\
			& MOST & 17793 & 18.43 \\\cline{2-4}
			& NSF & 33092 & 20.54  \\
			& NIH & 40353 & 25.05 \\\hline%
			\multirow{4}{*}{2011} & NSFC & 74294 & 63.16 \\
			& MOST & 20738 & 17.63 \\\cline{2-4}
			& NSF & 36077 & 20.65 \\ 
			& NIH & 42808 & 24.50 \\\hline%
			\multirow{4}{*}{2012} & NSFC & 93126 & 65.95 \\
			& MOST & 24274 & 17.19  \\\cline{2-4}
			& NSF & 35494 & 19.54  \\
			& NIH & 42025 & 23.13 \\\hline%
			\multirow{4}{*}{2013} & NSFC  & 117052 & 67.86 \\
			& MOST & 29972 & 17.37 \\\cline{2-4}
			& NSF & 34827 & 18.85 \\ 
			& NIH & 40838 & 22.11 \\\hline%
			\multirow{4}{*}{2014} & NSFC & 140599 & 69.46 \\
			& MOST & 33493 & 16.54 \\\cline{2-4}
			& NSF & 34129 & 18.46  \\
			& NIH & 38437 & 20.79 \\\hline%
			\multirow{4}{*}{2015} & NSFC & 161856 & 70.23 \\
			& MOST & 34701 & 15.05 \\\cline{2-4}
			& NSF & 35141 & 17.67  \\
			& NIH & 37382 & 18.80 \\\hline%
			\multirow{4}{*}{2016} & NSFC  & 180271 & 71.11 \\
			& MOST & 31972 & 12.61 \\\cline{2-4}
			& NSF & 35262 & 16.78  \\
			& NIH & 38026 & 18.09 \\\hline%
			\multirow{4}{*}{2017} & NSFC & 201188 & 71.63 \\
			& MOST & 32730 & 11.65 \\\cline{2-4}
			& NSF & 35170 & 16.24  \\
			& NIH & 39359 & 18.18 \\
			\hline
			
			\caption{The number of papers that acknowledge NSFC, MOST, NSF and NIH. The counting is based on whole counting method. The percentage is calculated as the ratio between the number of papers acknowledge the corresponding agency and the number of papers from China or the U.S.A in that year. We only count papers from China when counting NSFC and MOST supported papers, and take only papers from the U.S.A when dealing with NSF and NIH supported papers.} \label{table2}
		\end{longtable}
	\end{center}   
	\clearpage
	\begin{center}
		\renewcommand\arraystretch{0.8}		
		\begin{longtable}{|p{3cm}<{\centering}|p{4cm}<{\centering}|p{4.5cm}<{\centering}|}
			\hline
			Year & Funding Agency & Percentage of Overlap(\%) \\
			\hline%
			\multirow{4}{*}{2008} & NSFC & 16.43 \\
			& MOST & 65.87 \\\cline{2-3}
			& NSF & 11.40 \\
			& NIH & 8.32  \\\hline
			\multirow{4}{*}{2009} & NSFC & 18.90 \\
			& MOST & 68.45 \\\cline{2-3}
			& NSF & 10.98 \\
			& NIH & 8.18 \\\hline%
			\multirow{4}{*}{2010} & NSFC & 21.19 \\
			& MOST & 70.50 \\\cline{2-3}
			& NSF & 10.76 \\
			& NIH & 8.82 \\\hline%
			\multirow{4}{*}{2011} & NSFC & 19.98 \\
			& MOST & 71.59 \\\cline{2-3}
			& NSF & 11.09 \\
			& NIH & 9.34 \\\hline%
			\multirow{4}{*}{2012} & NSFC & 19.04 \\
			& MOST & 73.05  \\\cline{2-3}
			& NSF & 10.56 \\
			& NIH & 8.92 \\\hline%
			\multirow{4}{*}{2013} & NSFC & 19.24 \\
			& MOST & 75.14 \\\cline{2-3}
			& NSF & 10.08 \\
			& NIH & 8.60 \\\hline%
			\multirow{4}{*}{2014} & NSFC & 18.08 \\
			& MOST & 75.91 \\\cline{2-3}
			& NSF & 10.05 \\
			& NIH & 8.92 \\\hline%
			\multirow{4}{*}{2015} & NSFC & 16.56 \\
			& MOST & 77.26 \\\cline{2-3}
			& NSF & 9.68 \\
			& NIH & 9.10 \\\hline%
			\multirow{4}{*}{2016} & NSFC &13.79 \\
			& MOST & 77.77 \\\cline{2-3}
			& NSF & 10.13 \\
			& NIH & 9.40 \\\hline%
			\multirow{4}{*}{2017} & NSFC & 12.64 \\
			& MOST & 77.74 \\\cline{2-3}
			& NSF & 10.27 \\
			& NIH & 9.17 \\
			\hline%
			\caption{The percentage of overlap for NSFC is the ratio between the number of papers simultaneously supported by both NSFC and MOST in a particular year, and the number of papers supported by NSFC in that year. The percentage of overlap for MOST is the ratio between the number of papers simultaneously supported by both NSFC and MOST in a particular year, and the number of papers supported by MOST in that year. The same calculation is conducted for NSF and NIH. We only count papers from China when counting NSFC and MOST supported papers, and take only papers from the U.S.A when dealing with NSF and NIH supported papers. }
			 \label{table3}
		\end{longtable}
	\end{center}   
	
\end{document}